\documentclass[12pt]{article}

\usepackage{makeidx}
\usepackage{fix-cm}
\usepackage[utf8]{inputenc} 
\usepackage{times}
\usepackage{bm}
\usepackage{comment}
\usepackage{amssymb}
\usepackage{amsmath}
\usepackage{dirtytalk}
\usepackage{pdflscape}
\usepackage{listings}
\numberwithin{equation}{section}
\usepackage{color}
\usepackage{graphicx}
\usepackage{rotating}
\usepackage{pdflscape}
\usepackage{epstopdf}
\usepackage[round,authoryear,comma]{natbib}
\usepackage{natbib,hyperref}
\defcitealias{aci}{Actuaries Climate Index, 2016}
\usepackage{tikz}
\usetikzlibrary{positioning}

\usepackage{setspace} 
\usepackage{url}
\usepackage{booktabs}
\usepackage{actuarialangle}
\usepackage{actuarialsymbol}
\usepackage{setspace}
\usepackage{physics}
\usepackage{subcaption}
\usepackage{xcolor}
\usepackage{multirow}
\usepackage{caption}
\usepackage{authblk}
\usepackage{upgreek}
\usepackage{csquotes} 
\usepackage{bbm}
\usepackage{tabularx}
\usepackage{float}
\newcolumntype{Y}{>{\centering\arraybackslash}X} 
\hypersetup{
	colorlinks   = true, 
	urlcolor     = blue, 
	linkcolor    = blue, 
	citecolor   = blue 
}

\definecolor{OliveGreen}{rgb}{0,0.6,0}
\definecolor{alizarin}{rgb}{0.82, 0.1, 0.26}

\providecommand{\U}[1]{\protect\rule{.1in}{.1in}}
\parskip=\medskipamount

\begin{document}

\title{Mortality Modeling and Forecasting with the Actuaries Climate Index}

\author[1]{Karim Barigou\thanks{karim.barigou@uclouvain.be}}
\author[2]{Melanie Patten\thanks{mpatten2@asu.edu}}
\author[3]{Kenneth Q. Zhou\thanks{Corresponding author: kenneth.zhou@uwaterloo.ca}}

\affil[1]{Institute of Statistics, Biostatistics and Actuarial Science (ISBA), Louvain Institute of Data Analysis and Modeling (LIDAM), UCLouvain, Louvain-la-Neuve, Belgium} 
\affil[2]{School of Mathematical and Statistical Sciences, Arizona State University, USA}
\affil[3]{Department of Statistics and Actuarial Science, University of Waterloo, Canada}

\date{Version: \today}

\maketitle

\vspace{-0.8cm}

\begin{abstract}
Climate change poses increasing challenges for mortality modeling and underscores the need to integrate climate-related variables into mortality forecasting. This study introduces a two-step approach that incorporates climate information from the Actuaries Climate Index (ACI) into mortality models. In the first step, we model region-specific seasonal mortality dynamics using the Lee-Carter model with SARIMA processes, a cosine-sine decomposition, and a cyclic spline-based function. In the second step, residual deviations from the baseline model are explained by ACI components using Generalized Linear Models, Generalized Additive Models, and Extreme Gradient Boosting. To further capture the dependence between mortality and climate, we develop a SARIMA-Copula forecasting approach linking mortality period effects with temperature extremes. Our results show that incorporating ACI components systematically enhances out-of-sample accuracy, underscoring the value of integrating climate-related variables into stochastic mortality modeling. The proposed framework offers actuaries and policymakers a practical tool for anticipating and managing climate-related mortality risks.

\bigskip

\textbf{Keywords:} Mortality modeling; Climate risk; Actuaries Climate Index; Copula; Machine learning

\bigskip

\end{abstract}

\section{Introduction}

Climate change is a pressing global challenge that poses serious threats for the global health and society. The WHO has alerted that the frequency and severity of extreme weather events have all increased in the world. In an insurance context, several regulatory authorities emphasized the importance of the management and mitigation of climate risks, and the inclusion of climate risks within solvency capital requirements. To help insurance companies manage climate risks, North-American actuarial organizations developed the Actuaries Climate Index (ACI), which records information from several important weather variables. Since then, several countries have started constructing their own versions of the ACI \citep{australia,unitedkingdom,turkey,garrido:hal-04491982}. However, despite its potential, the ACI has seen limited practical implementation in research efforts and the insurance industry at large. This research aims to address this gap by investigating the dependence between mortality and the ACI components. 

The use of the ACI in the actuarial literature and the study of the impact of environmental factors on mortality forecasting are quite limited. \cite{pan2022assessing} adopts the index, as well as the constituting variables of the ACI, to build statistical models for estimating the impacts of extreme weather events on crop yields. The authors found that although the self-constructed ACI index leads to a slightly worse fit due to noisier county-specific yield data, the predictive results are still reasonable. \cite{seklecka2017mortality} proposed an adaptation of the Lee-Carter (LC) model by including a temperature-related exogenous factor, which outperforms the LC model. \cite{li2022joint} proposed an Extreme Value Theory (EVT) approach to model the extremal dependence between death counts and temperature using the multivariate peaks-over-threshold approach. More recently, \cite{robben2024association} proposed a two-step weekly mortality to study the impact of short-term association between environmental factors and weekly mortality rates based on a region-specific, seasonal trend baseline and a machine learning algorithm to explain deviations from environmental data. Focusing on 20 European countries, they found that environmental features are beneficial in explaining excess mortality for southern regions of Europe. \cite{guibert2024impacts} proposed a two-step modeling approach combining a stochastic mortality model with a climate epidemiology model. More specifically, a dynamic lag non-linear model (DLNM) was used to estimate temperature-attributable deaths and the multi-population Li-Lee model \citep{li2005coherent} was then used to estimate non-temperature-related deaths. Moreover, \cite{robben2025granular3states} proposed a three-state regime-switching model to capture excess mortality both due to extreme temperature and extreme influenza outbreaks. Finally, \cite{robben2025penalized} developed a forecasting framework that extends the LC model with age- and region-specific seasonal effects and penalized distributed lag non-linear components that capture the delayed and non-linear effects of heat, cold, and influenza on mortality.

A major part of the literature on stochastic mortality modeling stems from the seminal work of \cite{lee1992modeling}. Numerous extensions have been proposed to enhance forecasting performance (see, e.g., \cite{renshaw2006cohort, cairns2006two, plat2009stochastic}). However, a key limitation of these models lies in their reliance on latent factors to capture future mortality trends, leaving the underlying drivers of these trends ambiguous. To address this limitation, we propose a novel two-step framework that incorporates observable climate-related variables from the ACI into mortality modeling. This aligns with previous efforts to integrate observable factors such as Gross Domestic Product (GDP), Consumer Price Index (CPI), and unemployment rates into mortality forecasting (see, e.g., \cite{hanewald2011explaining, niu2014trends, boonen2017modeling, ma2022longevity}).  In particular, our research is in the spirit of \cite{robben2024association} who proposed a machine learning approach to explain deviations from a seasonal weekly mortality baseline using features constructed from environmental data that capture anomalies and extreme events. 

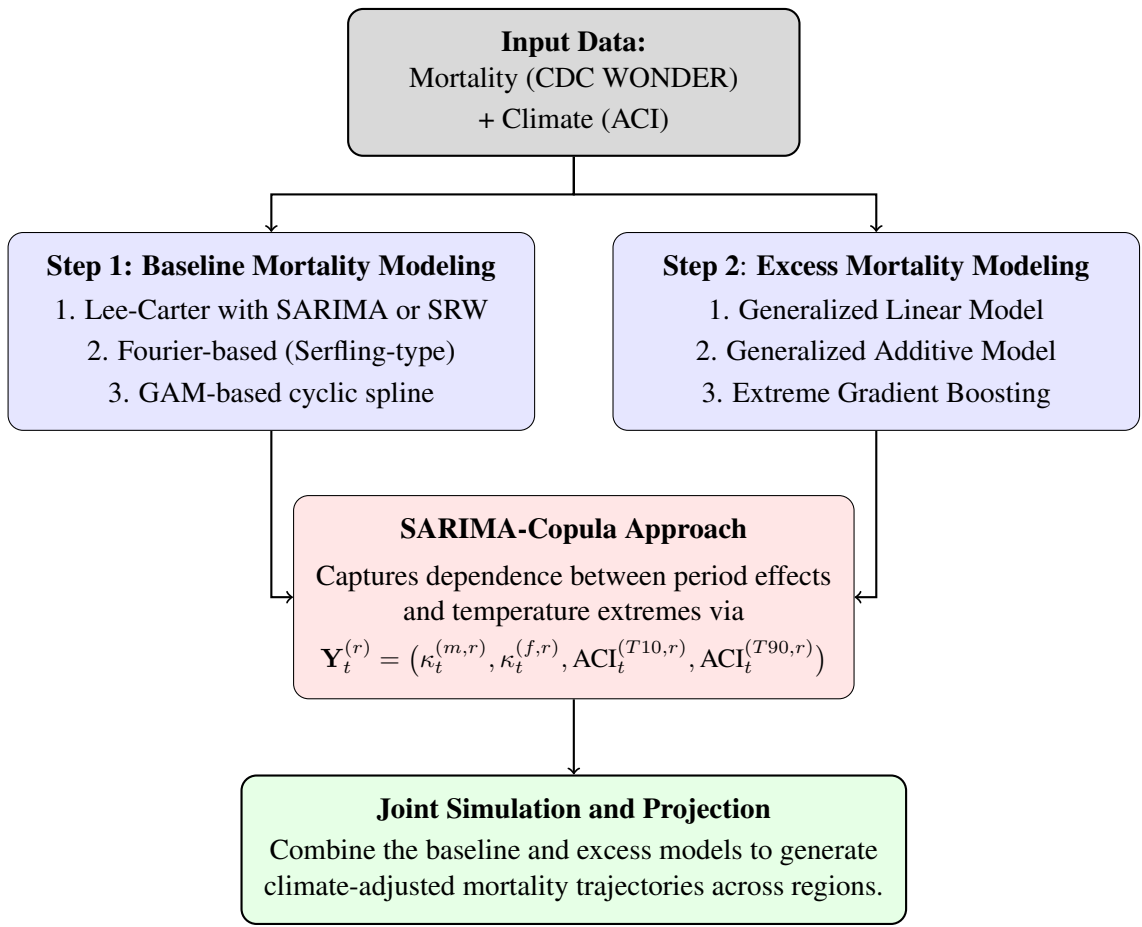
\begin{figure}[!ht]
\centering
\begin{tikzpicture}[
  node distance=1.3cm and 3.0cm,
  every node/.style={font=\small},
  box/.style={rectangle, draw, rounded corners=2mm, align=center, minimum width=4cm, fill=white, inner sep=0.3cm},
  bigbox/.style={rectangle, draw, rounded corners=2mm, align=center, minimum width=1cm, fill=blue!10, inner sep=0.3cm},
  integration/.style={rectangle, draw, rounded corners=2mm, align=center, minimum width=7cm, fill=red!10, inner sep=0.3cm},
  arrow/.style={->, thick}
]

\node[box, fill=gray!30, minimum width=6cm, thick] (data)
{ \textbf{Input Data:}\\ Mortality (CDC WONDER) \\[2pt] + Climate (ACI) };

\node[bigbox, below left=1cm and -2.5cm of data, anchor=north east] (step1) {
  \textbf{Step 1: Baseline Mortality Modeling}\\[6pt]
  \begin{minipage}{0.4\linewidth}
  \begin{center}
  1. Lee-Carter with SARIMA or SRW\\[2pt]
  2. Fourier-based (Serfling-type)\\[2pt]
  3. GAM-based cyclic spline
  \end{center}
  \end{minipage}
};

\node[bigbox, below right=1cm and -2.5cm of data, anchor=north west] (step2) {
  \textbf{Step 2}:
  \textbf{Excess Mortality Modeling}\\[6pt]
  \begin{minipage}{0.4\linewidth}
  \begin{center}
  1. Generalized Linear Model \\[2pt]
  2. Generalized Additive Model \\[2pt]
  3. Extreme Gradient Boosting
  \end{center}
  \end{minipage}
};

\node[integration, below=4.5cm of data] (copula) {
\textbf{SARIMA-Copula Approach}\\[4pt]
Captures dependence between period effects \\ and temperature extremes via\\[4pt]
{\footnotesize $\mathbf{Y}_t^{(r)} = \big(\kappa_t^{(m,r)}, \kappa_t^{(f,r)}, \text{ACI}_t^{(T10,r)}, \text{ACI}_t^{(T90,r)}\big)$}
};

\node[box, fill=green!10, minimum width=8cm, below=1cm of copula, thick] (forecast)
{\textbf{Joint Simulation and Projection}\\[2pt]
Combine the baseline and excess models to generate \\ climate-adjusted mortality trajectories across regions.};

\draw[arrow] (data.south) -- ++(0,-0.5) -| (step1.north);
\draw[arrow] (data.south) -- ++(0,-0.5) -| (step2.north);
\draw[arrow] (step1.south) |- (copula.west);
\draw[arrow] (step2.south) |- (copula.east);
\draw[arrow] (copula.south) -- (forecast.north);

\end{tikzpicture}
\caption{Schematic of the proposed mortality modeling and forecasting framework.}
\label{fig:two_step_framework}
\end{figure}

The proposed two-step monthly mortality modeling and forecasting framework is illustrated in Figure \ref{fig:two_step_framework}. The first step is the region-specific seasonal baseline mortality model. For this aim, we compare three baseline approaches: (1) a LC model with seasonal effects, incorporating either a seasonal random walk or SARIMA processes; (2) Serfling-type mortality model, incorporating sine and cosine Fourier terms to capture monthly seasonality \citep{serfling1963methods,robben2024association} or (3) a cyclic spline model. The second step aims to explain residuals from the baseline mortality model as a function of ACI components (high and low temperatures, heavy winds, heavy precipitation, and drought). To capture the relationship between these climate variables and residual mortality, we compare three statistical models, namely Generalized Linear Model (GLM), a Generalized Additive Model (GAM), and an Extreme Gradient Boosting (XGB) regression model. Finally, to capture and forecast seasonal dependencies between ACI components and mortality trends, we propose a SARIMA-Copula model. This integrated framework enables us to explore both the direct and residual impacts of climate variables on mortality dynamics.  

To evaluate the contribution of climate variables, we compare the performance of the three baseline approaches as well as the three models for explaining residuals, demonstrating the added value of including ACI components in out-of-sample forecasting. Additionally, we provide a detailed analysis of the dependence structure between temperature-related ACI components and the LC period effect $\kappa_t$ using a SARIMA-Copula framework, providing new insights into the interaction between climate variables and mortality dynamics. Our results indicate that incorporating climate-related variables into mortality modeling can significantly improve out-of-sample forecasting performance across all baselines. From an actuarial perspective, our work highlights the importance of linking climate risk with traditional mortality forecasting, providing practitioners and policymakers with enhanced modeling tools for solvency stress testing, experience studies, and risk management under changing climate conditions.

The paper is structured as follows. Section \ref{sec:Data} presents the data at hand. We introduce our general two-step modeling approach in Section \ref{sec:Model}. Section \ref{sec:estimation} discusses the estimation procedure. Section \ref{sec:results} discusses the comparison between the three baselines, the overall out-of-sample performance of the two-step modeling approach. Section \ref{sec:Forecast} presents the SARIMA-Copula approach for mortality forecasting. Section \ref{sec:conclusion} concludes. 

\section{Data description} \label{sec:Data}

Though our modeling approach is general and can be applied to any country with climate indexes, our empirical study will focus on U.S. monthly climate indexes and mortality data during 1999-2020. Hereafter, we briefly present the climate and mortality data. 

\subsection{Climate data} \label{ACIdescription}

To obtain climate-related data, one of the most notable actuarial indices will be utilized: the Actuaries Climate Index (ACI). The ACI, available on both a monthly and seasonal basis, is a collaborative monitoring tool developed by prominent actuarial organizations in North America, including the Canadian Institute of Actuaries, the Society of Actuaries, the Casualty Actuarial Society, and the American Academy of Actuaries (\citetalias{aci}). The index was created with the reference period of 1960-1990, and subsequent reporting is based on this standardized time frame.

The ACI comprises six components, each providing valuable insights into climate-related variations and extreme weather conditions:
(1) Frequency of temperatures above the 90th percentile (T90),
(2) Frequency of temperatures below the 10th percentile (T10),
(3) Maximum rainfall per month in five consecutive days (P),
(4) Annual maximum consecutive dry days (D),
(5) Frequency of wind speed above the 90th percentile (W), and
(6) Sea level changes (S).

\begin{figure}[hbt!]
    \centering
    \includegraphics[width=0.8\textwidth]{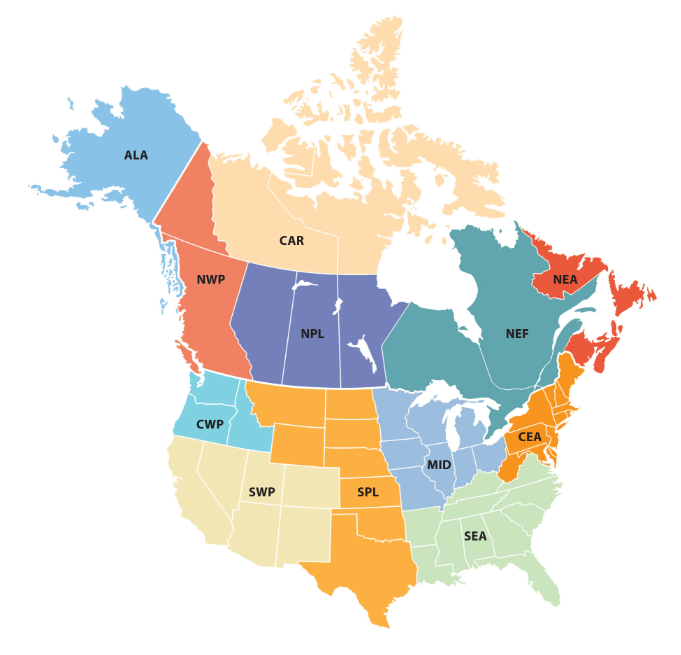}
    \caption{Regional groupings of the Actuaries Climate Index across North America. Source: Actuaries Climate Index Executive Summary (2018).}
    \label{ACI_Regions}
\end{figure}

These components are systematically constructed on a uniform grid, covering regions across the USA and Canada. In total, the ACI defines 12 distinct regions shown in Figure \ref{ACI_Regions}, enabling comprehensive monitoring and analysis of climate-related variables. Since our study focuses on the US, we use the monthly indices across USA regions in our analysis. However, we note that the Sea Level component of the index is not provided for the Midwest region (MID). Given its limited influence on overall mortality and the lack of data for this region, we exclude the Sea Level component from our analysis to ensure consistency across all regions.

\begin{figure}[hbt!]
    \centering
    \includegraphics[width=0.9\textwidth]{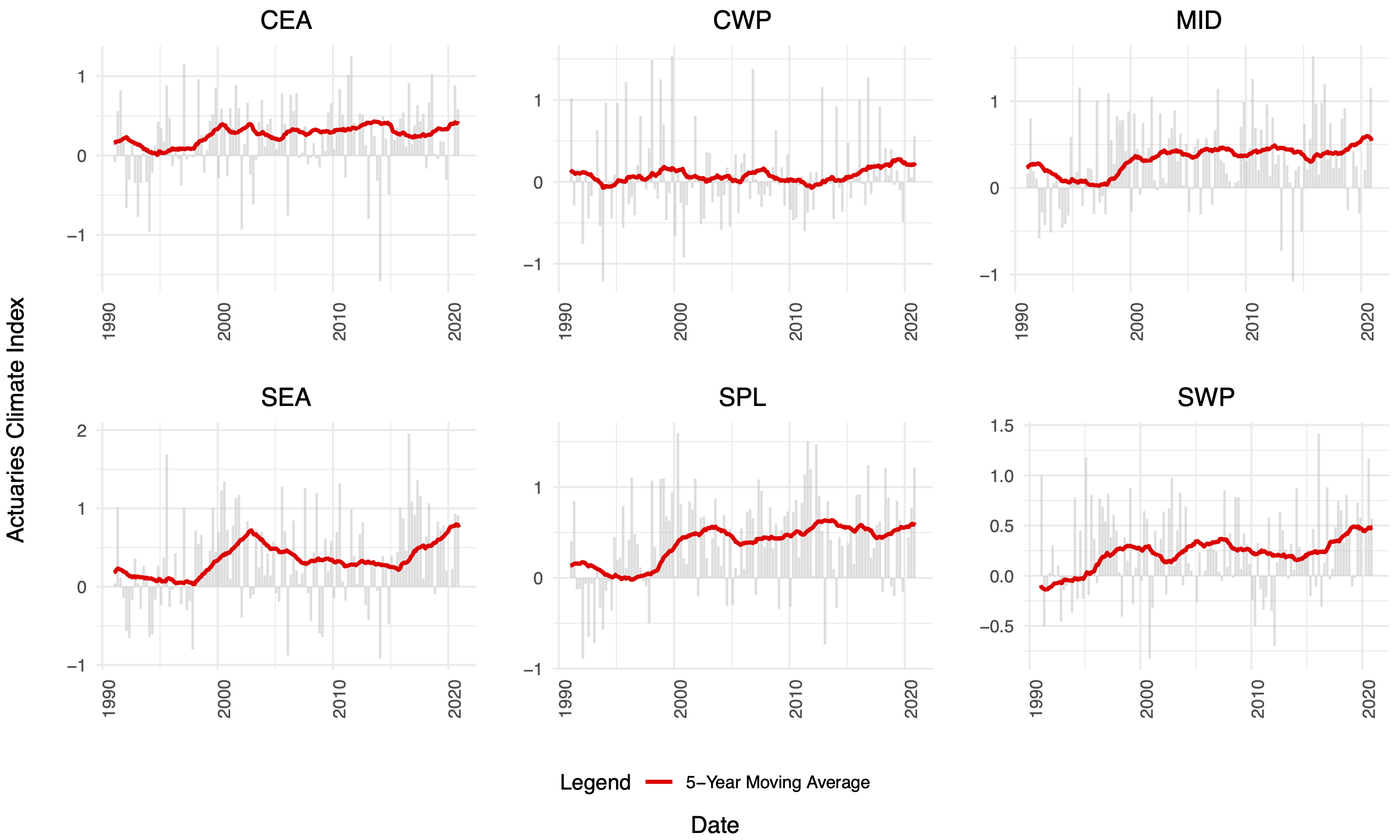}
    \caption{Standardized seasonal ACI values for six U.S. regions: Central East Atlantic (CEA), Central West Pacific (CWP), Midwest (MID), Southeast Atlantic (SEA), Southern Plains (SPL), and Southwest Pacific (SWP).}
    \label{ACI_unsmoothed_plots_by_region}
\end{figure}

While the North American actuarial organizations responsible for the ACI publish both a smoothed and an unsmoothed version of the components by month upon standardization, the raw unsmoothed version is preferred since the smoothing might reduce the observed impact of climate variables. For illustration, Figure \ref{ACI_unsmoothed_plots_by_region} displays the standardized seasonal averaged ACI by region. Across all regions, we observe a clear upward trend that shows the increase in extreme environmental events. 

\subsection{Mortality data} \label{MortalityData}

The USA actual mortality data over the investigation period (1999–2019) are gathered from the Centers for Disease Control and Prevention (CDC) WONDER, specifically the Underlying Cause of Death database. For the analysis, aggregated deaths by ACI region, gender, five-year age bands, and month were used, in order to match the monthly frequency of the regional ACI indices. Moreover, this study will focus on the 6 continental U.S. Regions, excluding the states of Alaska and Hawaii. Given the absence of publicly available monthly age-specific population exposures at the state level, we approximated these exposures by dividing yearly age-specific values by 12. While this approach assumes uniform exposure across months, it aligns with established practices in similar studies \citep{robben2022assessing}.

\section{Mortality modeling with ACI} \label{sec:Model}

The proposed monthly mortality model aims to explore the relationship between excess mortality and environment data, using the components of the ACI. For this purpose, we employ a two-step modeling approach. First, in Section \ref{section:baseline}, we fit a baseline monthly mortality model to capture underlying trends and seasonality. We consider three approaches for the baseline model: (1) a seasonal LC model with the period effect \(\kappa_t\) modeled using either a seasonal random walk or SARIMA process, (2) a seasonal cos-sin function model in the spirit of \cite{robben2024association}, and (3) a cyclic spline GAM model.

In the second step in Section \ref{section:twostep}, we analyze the residuals from the baseline model by applying Generalized Linear Models (GLM), Generalized Additive Models (GAM), and Extreme Gradient Boosting (XGBoost) to link them with ACI components. This approach allows us to quantify the impact of climate factors on excess mortality, leveraging in particular the non-linear modeling capabilities of GAM and XGBoost.

\subsection{Modeling baseline mortality}\label{section:baseline}

We first start by establishing a baseline monthly mortality model to capture the underlying trends and seasonality before analyzing the residuals in relation to the ACI. For this first step, we model death counts as a Poisson regression model with population exposures as an offset as is standard practice in mortality modeling \citep{currie2016fitting,andres2018stmomo}. We assume that the observed monthly death counts during month $t$ for region $r$, gender $g$ and at age $x$ are realizations from a Poisson distributed random variable $D_{x,t}^{(r,g)}$:
\begin{equation*}
D_{x,t}^{(r,g)} \sim \operatorname{Poisson}\left(E_{x,t}^{(r,g)}\mu_{x,t}^{(r,g)}\right)
\end{equation*}
with $E_{x,t}^{(r,g)}$ and $\mu_{x,t}^{(r,g)}$, the corresponding central exposure and force of mortality, respectively. 

We consider three distinct modeling structures for $\mu_{x,t}^{(r,g)}$ described below.

\subsubsection{Baseline 1: Seasonal Lee-Carter model}

The first approach considers the Lee-Carter (LC) model \citep{lee1992modeling} applied separately by region \(r\) and gender \(g\). The mortality rate \(\mu_{x,t}^{(r,g)}\) for age \(x\) at time \(t\) is modeled as:

\begin{equation} \label{eq:LCmodel}
\log \mu_{x,t}^{(r,g)}=\alpha_{x}^{(r,g)}+\beta_{x}^{(r,g)} \kappa_{t}^{(r,g)},
\end{equation}
where \(\alpha_{x}^{(r,g)}\) is the intercept and \(\beta_{x}^{(r,g)}\) is the sensitivity to the period effect that are age-region-gender specific, and \(\kappa_{t}^{(r,g)}\) represents the period effect, capturing the overall mortality trend over time by gender and region. 

To ensure that the \(\kappa_t^{(r,g)}\) does not inadvertently capture the effects of mortality shocks, such as peaks of excess mortality, we consider a fitted version of \(\kappa_t^{(r,g)}\) that accounts for both seasonal patterns and underlying trends. This adjustment is crucial because directly using the raw \(\kappa_t^{(r,g)}\) may remove environmental-related shocks from the residuals, potentially reducing the effectiveness of the second step of our modeling framework. By smoothing \(\kappa_t^{(r,g)}\) through appropriate time series techniques, we aim to isolate the structural mortality trend and seasonal variations, ensuring that the residuals accurately reflect deviations due to climate or other unexplained factors. 

The usual approach for modeling the fitted \( \kappa_t^{(r,g)} \) involves applying a random walk or an ARIMA process to yearly data \citep{cairns2006two,andres2018stmomo,barigou2022bayesian}. However, given that our data is monthly and exhibits seasonal variations with higher mortality in winter and lower mortality in summer, we adapt these methods to account for the monthly periodicity. Specifically, we consider seasonal equivalents of these models, namely the seasonal random walk and the Seasonal AutoRegressive Integrated Moving Average (SARIMA) model:

\paragraph{Seasonal Random Walk:}

In this approach, the period effect \(\kappa_{t}^{(r,g)}\) is modeled as a seasonal random walk:

\begin{equation}
\kappa_{t}^{(r,g)} = \kappa_{t-12}^{(r,g)} + \mu^{(r,g)}+\epsilon_t,
\end{equation}
where \(\epsilon_t\) is a random error term with mean zero and variance \(\sigma^2\), representing random fluctuations around the seasonal pattern. This method captures the recurring annual seasonality in mortality rates while allowing the seasonal component to evolve over time.  

\paragraph{Seasonal ARIMA (SARIMA):}

Alternatively, the fitted \(\kappa_t^{(r,g)}\) can be modeled using a Seasonal ARIMA (SARIMA) process. The general SARIMA formulation is:  

\begin{equation}
\Phi_p(B)\Phi_P(B^s)(1-B)^d(1-B^s)^D \kappa_{t}^{(r,g)} = \Theta_q(B)\Theta_Q(B^s)\epsilon_t,\label{sarima}
\end{equation}
where \(B\) is the backward shift operator; \(p, q, P, Q, d, D\) define the orders of the AR, MA, and differencing components; and \(s\) represents the seasonal cycle (e.g., 12 for monthly data). This formulation captures both short-term dependencies and long-term trends, as well as seasonal patterns in mortality.  

By using the fitted $\kappa_t^{(r,g)}$ from one of these time-series models, we reduce the likelihood that the period effect directly absorbs short-term mortality shocks. As an alternative, one could apply robust cubic smoothing splines to the period effect, as discussed in \cite{robben2022assessing}. Another option is to model a yearly period effect $\kappa_t^{(r,g)}$ and include an additional seasonal component, following the approach of \cite{robben2025penalized}.

\subsubsection{Baseline 2: Fourier-Based seasonal modeling}

For the second baseline, we model the seasonal mortality behaviour using Fourier terms (see e.g. \cite{serfling1963methods,robben2024association}). More specifically, we assume that the monthly mortality has the following structure:
\begin{equation}
\log \mu_{x,t}^{(r,g)}=\beta_{0,x}^{(r,g)}+\beta_{1,x}^{(r,g)} t+\beta_{2,x}^{(r,g)} \sin \left(\frac{2 \pi t}{12}\right)+\beta_{3,x}^{(r,g)} \cos \left(\frac{2 \pi t}{12}\right),
\end{equation}
where \(\beta_{0,x}^{(r,g)}\) is the intercept, \(\beta_{1,x}^{(r,g)}\) represents the linear trend over time, and \(\beta_{2,x}^{(r,g)}\) and \(\beta_{3,x}^{(r,g)}\) capture the seasonal fluctuations through sine and cosine terms, respectively. Note that this baseline is gender-, region- and age-specific. Notice that in \cite{robben2024association}, the baseline was only region-specific.

\subsubsection{Baseline 3: Cyclic spline modeling}

The third approach involves a Generalized Additive Model (GAM) using cyclic splines to model seasonality in mortality. The mortality rate \(\mu_{x,t}^{(r,g)}\) is modeled as
\begin{equation}
\log(\mu_{x,t}^{(r,g)}) = \alpha_{x}^{(r,g)} + f_{x}^{(r,g)}(\text{month}) + g_{x}^{(r,g)}(\text{year}),\label{cyclic}
\end{equation}
where \(\alpha_{x}^{(r,g)}\) represents the intercept specific to each combination of age, gender, and region. The function \(f_{x}^{(r,g)}(\text{month})\) is a cyclic spline over the months, designed to capture the seasonal pattern, while \(g_{x}^{(r,g)}(\text{year})\) is a cyclic spline modeling the trend over years. Note that for each time index $t$ in \eqref{cyclic}, there corresponds one specific month (month$=1,2,\dots,12$) and one specific year (year$=1999,2000,\dots,2019$), i.e. $t=(\text{month},\text{year})$.

The cyclic spline function \(f_{x}^{(r,g)}(\text{month})\) is defined as
\[
f_{x}^{(r,g)}(\text{month}) = \sum_{k=1}^{K} \beta_{x,k}^{(r,g)} B^{(k)}(\text{month}),
\]
where \(B^{(k)}(\text{month})\) are the cyclic spline basis functions over the 12 months, and \(\beta_{x,k}^{(r,g)}\) are the coefficients specific to each age, gender, and region. 

Similarly, the cubic spline function \(g_{x}^{(r,g)}(\text{year})\) is expressed as:
\[
g_{x}^{(r,g)}(\text{year}) = \sum_{j=1}^{J} \gamma_{x,j}^{(r,g)} S^{(j)}(\text{year}),
\]
where \(S^{(j)}(\text{Year})\) are the cubic spline basis functions used to model the smooth trend over the years, and \(\gamma_{x,j}^{(r,g)}\) are the corresponding coefficients. The number of basis functions \(K\) and \(J\) are determined based on the data and the desired degree of smoothness selected using cross-validation \citep{wood2017generalized}.

\begin{figure}[th!]
    \centering
    \begin{subfigure}[b]{0.49\textwidth}
        \centering
        \includegraphics[width=\textwidth]{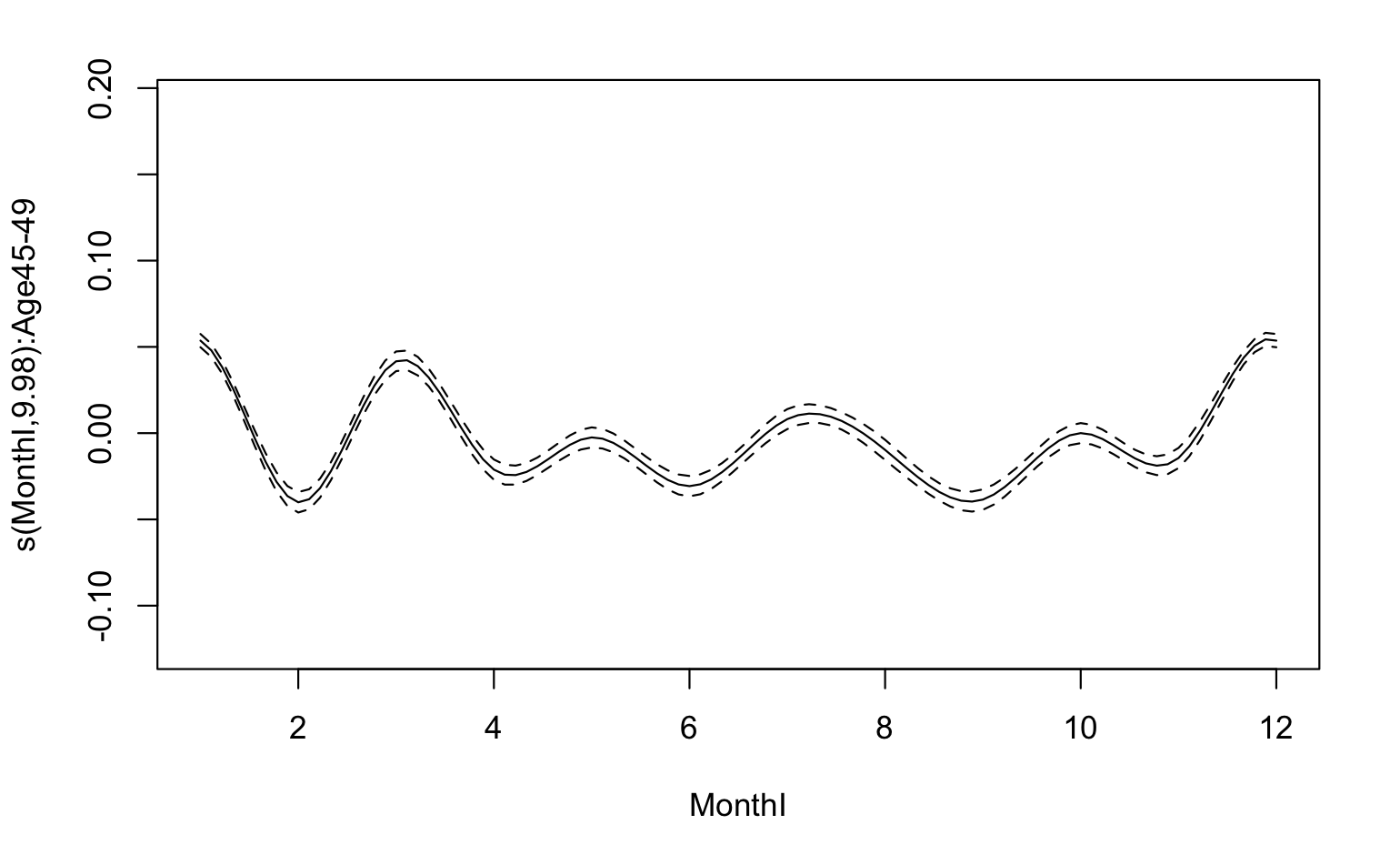}
        \caption{Monthly spline for ages 45-49}
        \label{fig:subfig1}
    \end{subfigure}
    \hfill
    \begin{subfigure}[b]{0.49\textwidth}
        \centering
        \includegraphics[width=\textwidth]{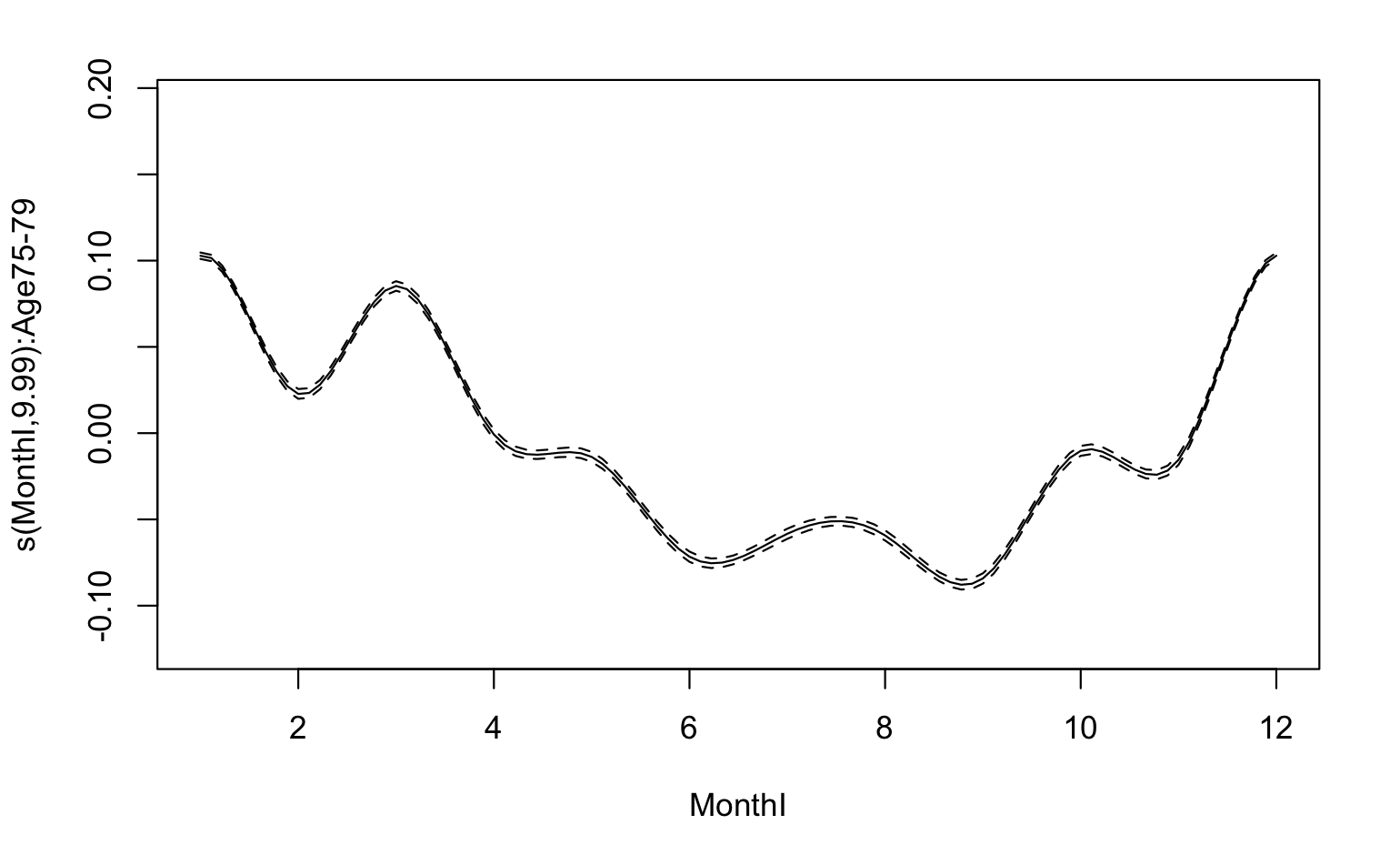}
        \caption{Monthly spline for ages 75-79}
        \label{fig:subfig2}
    \end{subfigure}
    \begin{subfigure}[b]{0.49\textwidth}
        \centering
        \includegraphics[width=\textwidth]{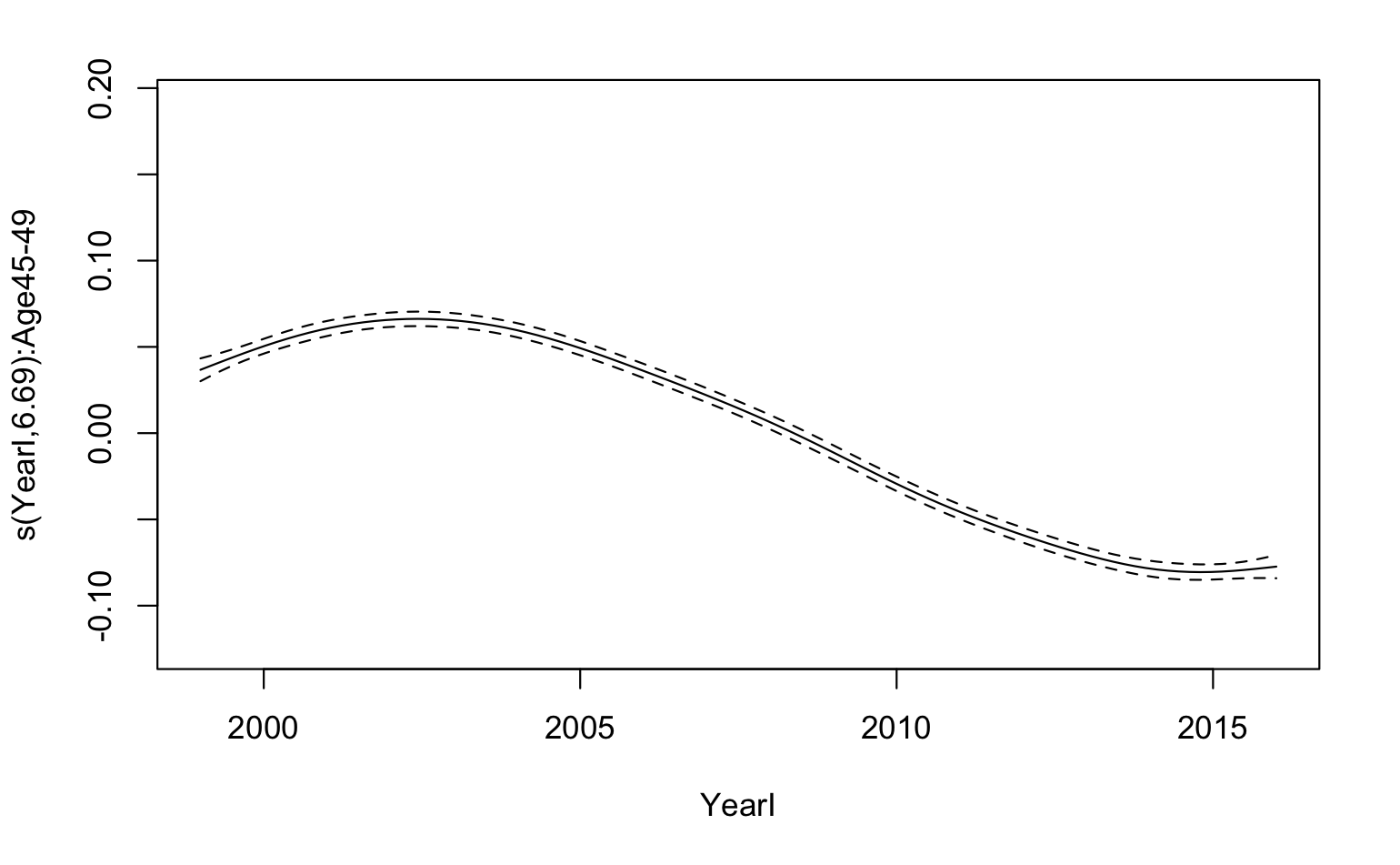}
        \caption{Yearly spline for ages 45-49}
        \label{fig:subfig3}
    \end{subfigure}
    \hfill
    \begin{subfigure}[b]{0.49\textwidth}
        \centering
        \includegraphics[width=\textwidth]{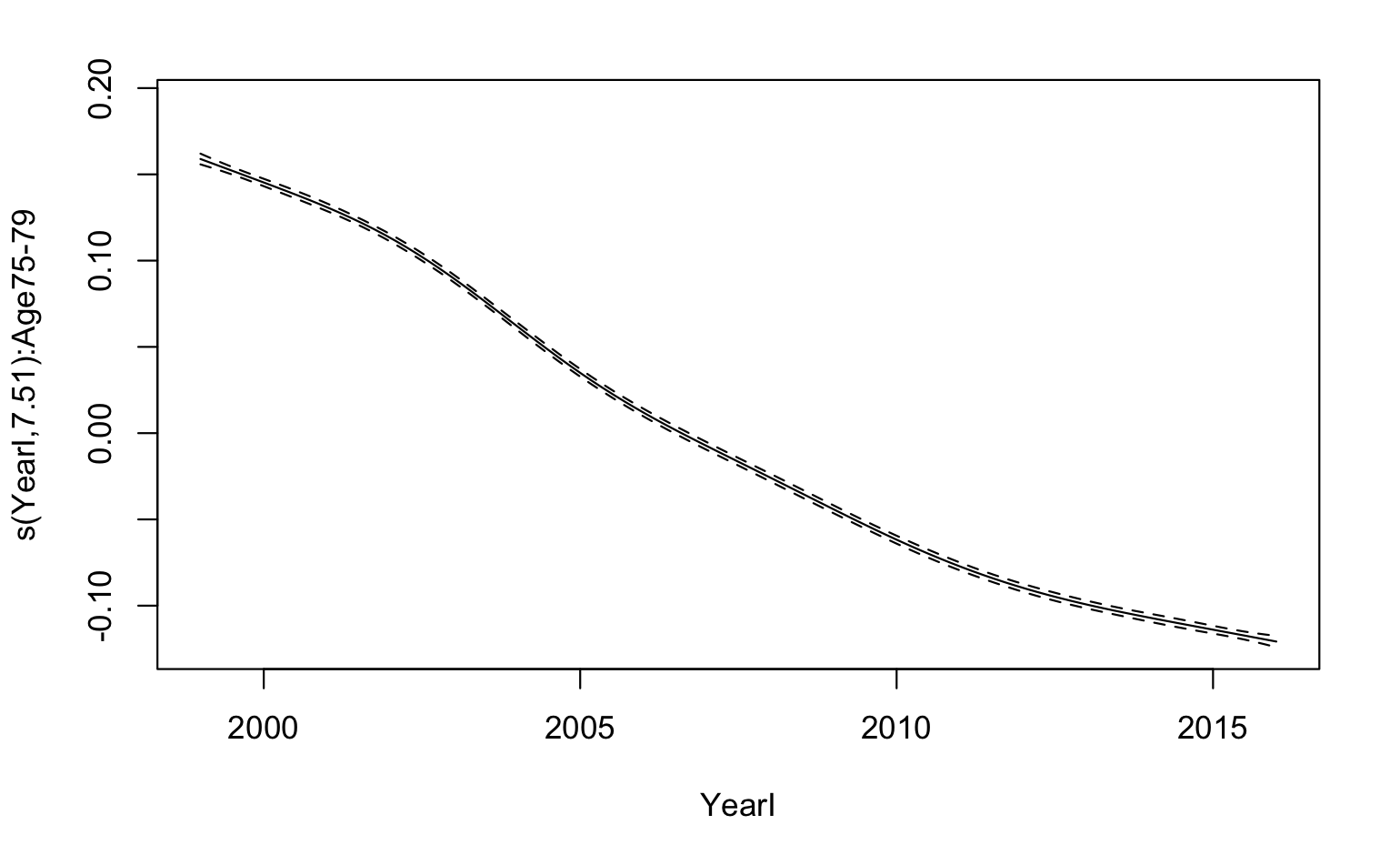}
        \caption{Yearly spline for ages 75-79}
        \label{fig:subfig4}
    \end{subfigure}
    \caption{Fitted monthly cyclic splines and yearly cubic splines for age groups 45-49 and 75-79.}
    \label{fig:MonthYearSplines}
\end{figure}

Figure \ref{fig:MonthYearSplines} shows the fitted spline functions of $f_{x}$ and $g_{x}$ for age groups $x=45-49$ and $x=75-79$. The fitted monthly splines shows a clear cyclic pattern over the 12 months, while the fitted year splines shows a clear downward trend over the sample years. Between the two age groups, it is apparent that the monthly and yearly splines are different, indicating that the young and old ages are facing different monthly and yearly mortality experiences.

\subsection{Modeling excess mortality}\label{section:twostep}

After fitting the baseline monthly mortality model, we aim to analyze the residuals to capture the relationship between excess mortality and the ACI components. To analyze excess mortality, we define the residuals, referred to as log-scale mortality residuals, as the difference between the observed log death rates and the estimated log death rates from the baseline model., that is
\[
R_{x,t}^{(r,g)} = \log \left(\frac{D_{x,t}^{(r,g)}}{E_{x,t}^{(r,g)}}\right) - \log \hat{\mu}_{x,t}^{(r,g)},
\]
where \( \hat{\mu}_{x,t}^{(r,g)} \) is the estimated mortality rate from the baseline model described in Section \ref{section:baseline}. Unlike deviance residuals from the Poisson regression, which are commonly used for model diagnostics (see e.g. \cite{andres2018stmomo}), these residuals are chosen for their ease of interpretation as direct deviations in log mortality rates. These residuals represent the excess or unexplained mortality, after accounting for the baseline trends and seasonality.

We model the residuals as a function of age, time, and the ACI components. Depending on the model used, we can capture linear or non-linear relationships between these covariates and the residuals. Specifically, we consider three approaches: Generalized Linear Models (GLM), Generalized Additive Models (GAM), and Extreme Gradient Boosting (XGBoost).

Unlike GLM, both GAM and XGBoost are adept at capturing non-linear relationships between covariates and the response variable. This capability is particularly pertinent, given the established non-linear dynamics between temperature-related climate variables and mortality. The environmental literature extensively documents the association between extreme temperatures---ranging from severe heatwaves to intense cold spells---and mortality rates \citep{baccini2008heat,ishigami2008ecological,hajat2010heat,yang2021projecting,xing2022projections}. Characteristically, the temperature-mortality relationship assumes a non-linear, U-shaped curve, where the extremes exert the most significant impact. Numerous studies have sought to quantify temperature's health effects by positing a linear relationship only beyond certain temperature thresholds \citep{baccini2008heat,mcmichael2008international}. Typically, these analyses incorporate a heat threshold---denoting the onset of adverse effects---and a heat slope, which quantifies the impact magnitude for each incremental degree above this threshold \citep{hajat2010heat}. Given this context, GAM and XGBoost are anticipated to outperform GLM, due to their superior capacity for modeling the nuanced, non-linear effects of climate variables on mortality. In the actuarial literature, \cite{robben2024association} considered the XGBoost machine learning approach to explain excess mortality and \cite{guibert2024impacts,robben2025penalized} used the DLNM model to capture deaths attributable to temperature variations.

\subsubsection{Explaining residuals via GLM}

In the GLM framework, the residuals are modeled as:

\[
\log \left(R_{x, t}^{(r, g)}\right) = \beta_0 + \sum_{a} \beta_{1a} \mathbb{I}(x = a) + \gamma t + \sum_k \delta_k \mathbf{ACI}_{t}^{(k,r)} + \epsilon_{x, t}^{(r, g)},
\]
where \( \beta_0 \) is the intercept, \( \sum_{a} \beta_{1a} \mathbb{I}(x = a) \) accounts for categorical age groups \( x \), \( \gamma \) is the coefficient for the linear time trend \( t \), and \( \delta_k \) are the coefficients for the ACI components \( \mathbf{ACI}_{t}^{(k,r)} \) specific to each region $r$. The error term \( \epsilon_{x, t}^{(r, g)} \sim N(0, \sigma_e^2) \).

Age (\( x \)) consists of 5-year groups (\([40,44], \ldots, [80,84]\)), and the analysis spans monthly time periods (\( t \): January 1999 to December 2019).

\subsubsection{Explaining residuals via GAM}
In the GAM framework, we extend the GLM by allowing for non-linear relationships between the covariates and the residuals. The GAM models the residuals as

\[
\log \left(R_{x,t}^{(r,g)}\right) = \beta_0  + f_1(x) + f_2(t) + \sum_k f_k(\text{ACI}_{t}^{(k,r)}) + \epsilon_{x,t}^{(r,g)},
\]
where \( f_i(\cdot) \) represents smooth functions that model non-linear effects for age, time trend, and each ACI component. These smooth functions allow for greater flexibility in capturing complex, non-linear relationships, particularly between the log residuals and the ACI components.

\subsubsection{Explaining residuals via XGBoost}

XGBoost, which stands for Extreme Gradient Boosting, is a leading algorithm within the gradient boosting framework. Introduced by \cite{chen2016xgboost}, it leverages decision trees as weak learners in an ensemble method, where each tree incrementally corrects the residuals of the ensemble of previously built trees. This iterative process improves the model's accuracy by focusing on the patterns not captured by previous trees.

The XGBoost algorithm models the residuals as:

\begin{equation}
   R_{x,t}^{(r,g)} = \sum_{k=1}^{K} f_k(x, t, \text{ACI}_{t}^{(k,r)}), \quad f_k \in \mathcal{F},
\end{equation}
where \(f_k\) represents an individual decision tree that operates on the extended feature space, and \(\mathcal{F}\) denotes the space of all possible decision trees. Each tree \(f_k\) is trained to minimize the objective function by incrementally reducing the residual error from the previous trees, allowing XGBoost to effectively capture complex, non-linear interactions among the demographic factors (age in this case), time trend, and ACI components.

The objective function optimized by XGBoost incorporates both the training loss and a regularization term to control model complexity. It is given by:

\begin{equation}
    \text{Objective} = \sum_{x, t, r, g} \ell(R_{x,t}^{(r,g)}, \hat{R}_{x,t}^{(r,g, t)}) + \sum_{k=1}^{K} \Omega(f_k),
\end{equation}
where \(\ell(R_{x,t}^{(r,g)}, \hat{R}_{x,t}^{(r,g, t)})\) is the loss function\footnote{In this case, we use the standard squared error loss function.} that compares the predicted residuals \(\hat{R}_{x,t}^{(r,g, t)}\) and the actual residuals \(R_{x,t}^{(r,g)}\). The term \(\Omega(f_k)\) denotes the regularization term for the \(k\)-th tree \(f_k\), which controls the complexity of the trees and helps mitigate overfitting by penalizing trees with more leaves or larger weights.

For the three model specifications (GLM, GAM and XGBoost), we consider three groups of covariates:
\begin{enumerate}
    \item \textbf{ACIs only}: This includes only the five climate-related indices (Heat Stress, Cold Stress, Heavy Precipitation, Drought, and Tropical Storms) without considering any demographic variables like age or time.
    \item \textbf{ACIs + Age}: This includes the five ACIs along with the age covariate $x$.
    \item \textbf{ACIs + Age + Time}: This specification includes the ACIs, age covariate $x$ and time covariate $t$.
\end{enumerate}

Even though age and time effects are already present in the baseline model discussed in Section \ref{section:baseline}, we find that adding age and time as covariates to explain excess mortality helps to improve out-of-sample performance as it will be shown later in the results section.

\section{Model estimation and results}\label{sec:estimation}

To assess the performance of our models, we utilize monthly data spanning from January 1999 to December 2016 as the in-sample data, and monthly data from January 2017 to December 2019 as the out-of-sample data. This temporal split allows for model training on historical data and validation on subsequent data to assess predictive accuracy.

\subsection{Estimation method}

To estimate the parameters, various methodologies can be employed. The conventional approach, as utilized in \cite{lee1992modeling}, involves the application of singular value decomposition (SVD) for parameter estimation in the LC model. An alternative method is the explicit estimation of parameters using ordinary least squares (OLS), as explored by \cite{liu2019statistical}. A third method involves the estimation of a generalized non-linear model, as proposed by \cite{currie2016fitting}. In this study, we adopt the SVD approach for each gender and region combination when fitting the LC model. Additionally, we employ the Hyndman-Khandakar algorithm \citep{hyndman2008automatic} to estimate the seasonal random walk and the SARIMA processes, utilizing the \texttt{forecast} R package. Generalized linear and additive models are estimated by performing standard log-likelihood maximization. 

To optimize the hyperparameters of the XGBoost algorithm, we employ 5-fold cross-validation. Specifically, we focus on tuning the following parameters: the regularization parameter \texttt{alpha}, the maximum depth of trees \texttt{max\_depth}, the learning rate \texttt{eta}, as well as the subsample fraction \texttt{subsample} and the fraction of features used for each split \texttt{colsample}. The cross-validation procedure helps to identify the set of hyperparameters that minimizes out-of-sample prediction error, thereby enhancing the model's performance. The optimal tuning parameters correspond to the parameter combination that yields the smallest average log-likelihood on the hold-out folds.

\subsection{Performance measurements}

To evaluate the out-of-sample performance of the models, we use three standard metrics based on \(\log(\mu_{x,t}^{(r,g)})\): Root Mean Squared Error (RMSE), Mean Absolute Percentage Error (MAPE), and Mean Absolute Error (MAE).

\begin{itemize}
    \item \textbf{Root Mean Squared Error (RMSE):} RMSE measures the square root of the average squared difference between the observed and predicted log mortality rates. It is defined as:
    \[
    \text{RMSE} = \sqrt{\frac{1}{N} \sum_{x, t, r, g} \left(\log(\mu_{x,t}^{(r,g)}) - \log(\hat{\mu}_{x,t}^{(r,g)})\right)^2},
    \]
    where \(\log(\mu_{x,t}^{(r,g)})\) are the observed log mortality rates, \(\log(\hat{\mu}_{x,t}^{(r,g)})\) are the predicted log mortality rates from the estimated model, and \(N\) is the number of observations. The RMSE provides a measure of the average magnitude of error in the model’s predictions, with the advantage of being in the same units as the log mortality rates.

    \item \textbf{Mean Absolute Percentage Error (MAPE):} MAPE measures the average percentage error between the observed and predicted log mortality rates. It is defined as:
    \[
    \text{MAPE} = \frac{100\%}{N} \sum_{x, t, r, g} \left|\frac{\log(\mu_{x,t}^{(r,g)}) - \log(\hat{\mu}_{x,t}^{(r,g)})}{\log(\mu_{x,t}^{(r,g)})}\right|.
    \]

    \item \textbf{Mean Absolute Error (MAE):} MAE computes the average absolute difference between the observed and predicted log mortality rates. It is defined as:
    \[
    \text{MAE} = \frac{1}{N} \sum_{x, t, r, g} \left|\log(\mu_{x,t}^{(r,g)}) - \log(\hat{\mu}_{x,t}^{(r,g)})\right|.
    \]
\end{itemize}

\subsection{Model results} \label{sec:results}

This subsection is organized into two parts. First, we compare the performance of the three baseline models introduced in Section~\ref{section:baseline}, focusing on their ability to capture and forecast mortality patterns. Second, we evaluate the out-of-sample performance of our two-step modeling approach with different models for the residuals—GLM, GAM, and XGBoost—under varying covariate specifications: ACI only, ACI + Age, and ACI + Age + Time. This analysis sets the stage for incorporating these components into the SARIMA-Copula forecasting framework discussed in Section~\ref{sec:Forecast}.

\subsubsection{Baseline model results}

Table \ref{tab:BaselineMeasures} presents the RMSE, MAPE, and MAE metrics on the testing period (2017–2019) for the three baseline models introduced in Section \ref{section:baseline}. Baseline 1 corresponds to the LC model with either SARIMA or Seasonal Random Walk (SRW) to model the period effect. Among these, the LC-SARIMA model outperforms the LC-SRW model, demonstrating improved forecasting accuracy thanks to its greater flexibility to adjust the observed time series. We recall that Baseline 2 is a GLM based on sine and cosine functions of the monthly index with coefficients specific to age, gender and region. Although this model incorporates periodic trigonometric components, it exhibits the largest error metrics among all baselines, likely due to its limited flexibility in capturing complex mortality dynamics. 

\begin{table}[ht!]
    \centering
    \begin{tabular}{c|l|c|c|c}
        \hline \hline
        \textbf{Baseline} & \textbf{Model} & \textbf{RMSE} & \textbf{MAPE} & \textbf{MAE} \\ \hline
        1      & LC-SARIMA                 & 0.06876 & 0.01109 & 0.05136      \\ \hline
        1      & LC-SRW          & 0.07243 & 0.01173 & 0.05421      \\ \hline
        2      & GLM with sin-cos terms         & 0.09190 & 0.01633 & 0.07330      \\ \hline
        3      & GAM cyclic splines by age         & 0.07163 & 0.01138 & 0.05294      \\ \hline
        3      & GAM cyclic splines by age and gender    & 0.07059 & 0.01106 & 0.05163      \\ \hline
        3      & GAM cyclic splines by age and region    & 0.06739 & 0.01062 & 0.04915      \\ \hline
        3      & GAM cyclic splines by gender and region & 0.08063 & 0.01398 & 0.06327      \\ \hline
        3      & GAM cyclic splines by age, gender and region  &  0.06432 & 0.01005 & 0.04653      \\ \hline \hline
    \end{tabular}
    \caption{RMSE, MAPE and MAE metrics on the testing period for different baseline models.}
    \label{tab:BaselineMeasures}
\end{table}

Baseline 3 employs Generalized Additive Models (GAMs) with monthly and yearly cyclic cubic splines. When splines are fitted to different combinations of covariates, the performance varies. We note the following:
\begin{itemize}
    \item The GAM by age and region achieves superior performance across all error metrics compared to LC-SARIMA, while maintaining a balance between model complexity and accuracy.
    \item The GAM by age, gender, and region produces the best overall fit but at the cost of increased complexity. For practical purposes, the GAM by age and region strikes a favorable trade-off, making it a strong candidate for Baseline 3.
\end{itemize}

\begin{figure}[th!]
    \centering
    \includegraphics[width=\textwidth]{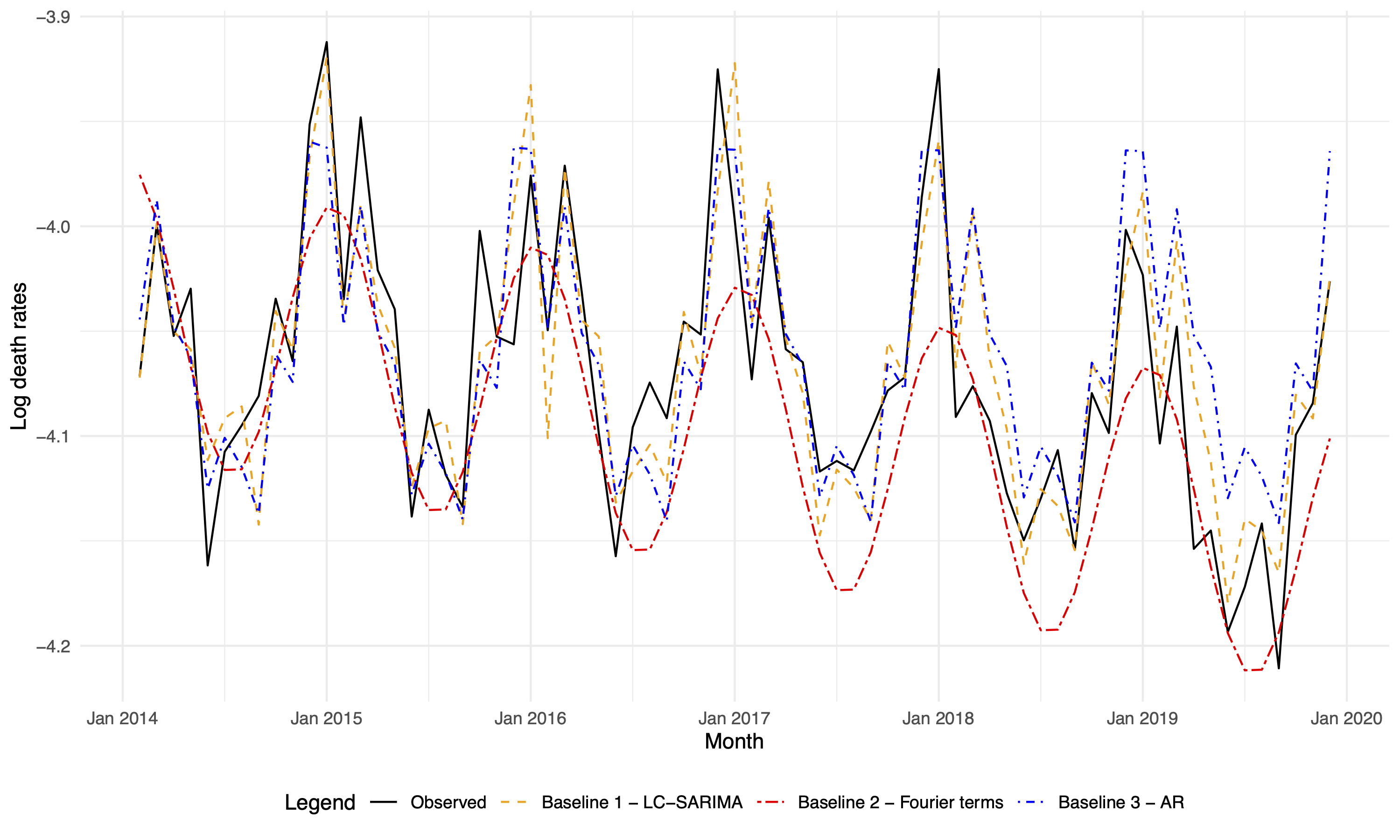}
    \caption{The observed and fitted log death rates from Baseline 1 (LC-SARIMA), Baseline 2, and Baseline 3 (Cyclic splines by age and region) from Jan 2014 to Dec 2019 (3 training years plus 3 testing years) for female, age group 70-74, from the CEA region.}
    \label{fig:BaselineCompare}
\end{figure}

Figure \ref{fig:BaselineCompare} illustrates the observed and fitted log death rates from Baseline 1 (LC-SARIMA), Baseline 2 (GLM with sine-cosine terms), and Baseline 3 (GAM by age and region) for the age group 60–64 in the CEA region over the period January 2014 to December 2019. The following observations can be drawn:
\begin{itemize}
    \item Baseline 1 and 3 closely follows the observed rates, both in the training (2014–2016) and testing (2017–2019) periods. Importantly, it successfully captures specific region- and age-related patterns.
    \item Baseline 2 (GLM with sine-cosine terms) shows more periodic deviations from the observed rates, particularly in the testing period, likely due to the rigid nature of its trigonometric components.
\end{itemize}

Following this, we select the SARIMA specification for Baseline 1 due to its strong performance, while GAM by age and region is chosen for Baseline 3, as it outperforms LC-SARIMA while remaining less complex than GAM by age, gender, and region.

\subsubsection{Two-step model results}

This subsection examines the best-performing model for each baseline and each evaluation metric (MAE, MAPE, RMSE), as well as the overall best model (GLM, GAM, XGBoost) across all three baselines. We also discuss the results by age classes and by US regions. By identifying the optimal configurations, we aim to assess the relative strengths of different modeling approaches and covariate selections, emphasizing their ability to capture the underlying mortality dynamics effectively.


The results in Table~\ref{tab:best_models} highlight the strong performance of XGBoost across both Baseline~1 (LC--SARIMA) and Baseline~2 (cosine--sine functions). Overall, Baseline~1 achieves the best performance for all evaluation metrics (MAE, MAPE, and RMSE) when the ACI variables are combined with age and time covariates. The table also shows that including age and time, in addition to the five ACI components, improves the out-of-sample performance. This improvement likely reflects that the influence of climatic factors on mortality varies by age group and over time, consistent with the findings of \cite{robben2025granular3states}.

\begin{table}[!ht]
\centering
\begin{tabular}{llccc}
\toprule
\textbf{Baseline} & \textbf{Measure} & \textbf{Model} & \textbf{Regressors} & \textbf{Value} \\
\midrule
Baseline 1 & MAE  & XGBoost & ACIs + Age + Time& 0.0467 \\ 
           & MAPE & XGBoost & ACIs + Age + Time& 0.0102 \\ 
           & RMSE & XGBoost & ACIs + Age + Time& 0.0628 \\ 
\midrule
Baseline 2 & MAE & XGBoost & ACIs + Age + Time & 0.0538 \\
           & MAPE & XGBoost & ACIs + Age + Time & 0.0117 \\ 
           & RMSE & XGBoost & ACIs + Age + Time & 0.0716 \\ 
\midrule
Baseline 3 & MAE & GAM & ACIs + Age & 0.0473 \\
           & MAPE & GAM & ACIs + Age & 0.0102 \\ 
           & RMSE & GAM & ACIs + Age & 0.0648 \\ 
\midrule
\textbf{Overall Best: Baseline 1} & \textbf{MAE}  & \textbf{XGBoost} & \textbf{ACIs + Age + Time } & \textbf{0.0467} \\
                      & \textbf{MAPE} & \textbf{XGBoost} & \textbf{ACIs + Age + Time} & \textbf{0.0102} \\
                      & \textbf{RMSE} & \textbf{XGBoost} & \textbf{ACIs + Age + Time} & \textbf{0.0628} \\
\bottomrule
\end{tabular}
\caption{Best model specifications (XGBoost, GAM, GLM) and regressors for each baseline and overall best model.}
\label{tab:best_models}
\end{table}

Baseline 2, while also using XGBoost, exhibits slightly inferior performance compared to Baseline 1, with higher error values across all measures. This difference suggests that the cos-sin baseline lacks of flexibility compared to the LC model with SARIMA time series. Baseline 3, which uses a GAM model with ACIs and age, remains competitive and achieves comparable results in terms of MAPE and RMSE. However, it falls short of the overall performance achieved by Baseline 1 with XGBoost but not by a great extent. The simplicity and interpretability of Baseline 3 may still make it an appealing choice for scenarios prioritizing model transparency over marginal gains in predictive accuracy. Overall, the results highlight XGBoost's effectiveness in handling complex interactions between demographic and climate-related variables.


The results in Table \ref{tab:comparison_by_age} reveal notable variations in model performance across age categories. Younger age groups, such as 40-44 and 55-59, show significant improvements when using XGBoost models incorporating ACI variables, age, and time indices, with RMSE reductions of 14.78\% and 18.07\%, compared to LC-SARIMA and LC-SRW, respectively. These improvements highlight the ability of advanced machine learning models to effectively capture nuanced patterns in mortality trends for these demographics. Conversely, older age groups, such as 75-79, exhibit smaller relative improvements, suggesting that mortality dynamics in these cohorts may be less sensitive to the included predictors or that additional covariates may be required to further enhance model performance.

\begin{table}[ht!]
\centering
\begin{tabular}{llccccc}
\hline
Age Category & \multicolumn{2}{c}{Model} & Regressors & RMSE & \multicolumn{2}{c}{Improvement (\%)} \\ 
\cline{2-3} \cline{6-7}
& Baseline & Residual & & & LC-SARIMA & LC-SRW \\ \hline
40-44 & Baseline 1 & XGB & ACIs + Age + Time & 0.1005 & 14.78 & 18.03 \\ 
45-49 & Baseline 3 & GAM & ACIs + Age & 0.0727 & 7.80 & 9.83 \\ 
50-54 & Baseline 1 & GLM & ACIs + Age & 0.0708 & 0.57 & 0.81 \\ 
55-59 & Baseline 1 & XGB & ACIs + Age + Time & 0.0590 & 18.07 & 20.62 \\ 
60-64 & Baseline 1 & XGB & ACIs + Age + Time & 0.0448 & 33.61 & 38.97 \\ 
65-69 & Baseline 1 & GAM & ACIs + Age + Time & 0.0442 & 5.99 & 24.72 \\ 
70-74 & Baseline 1 & GLM & ACIs & 0.0404 & 0.95 & 6.21 \\ 
75-79 & Baseline 1 & GLM & ACIs & 0.0408 & 0.69 & 8.40 \\ 
80-84 & Baseline 3 & XGB & ACIs + Age + Time & 0.0429 & 6.63 & 12.26 \\ \hline
\end{tabular}
\caption{Comparison of the best baseline and residual models by age classes.}
\label{tab:comparison_by_age}
\end{table}

Table \ref{tab:comparison_by_region} demonstrates regional variations in the effectiveness of the best-performing models. In regions such as MID, where the XGBoost model with ACIs, age, and time indices achieves a 27.81\% improvement in RMSE over LC-SARIMA, the integration of demographic and climate-related variables proves highly impactful. Conversely, regions like SWP show minimal improvement (0.11\%), indicating that the predictive power of the chosen covariates may vary across geographic areas. This variability underscores the importance of regional customization in model specifications, as different regions may require tailored approaches to address unique demographic and environmental factors influencing mortality dynamics.

\begin{table}[ht!]
\centering
\begin{tabular}{llccccc}
\hline
Region & \multicolumn{2}{c}{Model} & Regressors & RMSE & \multicolumn{2}{c}{Improvement (\%)} \\ 
\cline{2-3} \cline{6-7}
& Baseline & Residual & & & LC-SARIMA & LC-SRW \\ \hline
CEA & Baseline 1 & XGB & ACIs + Age + Time & 0.0517 & 18.19 & 22.49 \\ 
CWP & Baseline 1 & XGB & ACIs + Age + Time & 0.0941 & 5.49 & 7.49 \\ 
MID & Baseline 1 & XGB & ACIs + Age + Time & 0.0474 & 27.81 & 31.48 \\ 
SEA & Baseline 3 & GLM & ACIs + Age & 0.0525 & 11.65 & 16.11 \\ 
SPL & Baseline 3 & GAM & ACIs + Age + Time & 0.0571 & 6.53 & 13.67 \\ 
SWP & Baseline 1 & GAM & ACIs & 0.0540 & 0.11 & 10.03 \\ \hline
\end{tabular}
\caption{Comparison of the best baseline and residual models by region.}
\label{tab:comparison_by_region}
\end{table}

\section{Mortality forecasting with ACI} \label{sec:Forecast}

Up to this point, we have been focusing on fitting suitable models to capture the seasonality of monthly mortality trends and their dependence on the ACIs. In this section, we shift our attention to how mortality rates and ACIs can be \emph{jointly} modeled and projected into the future. The forecasting framework developed here combines marginal time-series dynamics with multivariate dependence, allowing us to generate realistic paths for both mortality period effects and climate-related indices.

Numerous studies have demonstrated that temperature extremes are among the most significant climate-related factors affecting mortality rates. Specifically, both low temperatures (T10) and high temperatures (T90) have been shown to have a profound impact on mortality \citep{robben2024association,guibert2024impacts,gasparrini2015mortality}. In particular, the feature importance in \cite{robben2024association} using XGBoost revealed that extreme temperatures are the most significant features among a large group of environment-related features. Cold temperatures can lead to hypothermia and exacerbate respiratory diseases, while high temperatures can cause heat stroke and worsen cardiovascular conditions. 

While the ACI includes six components, both the literature and our own results indicate that T10 and T90 are the dominant drivers of mortality risk. Given this evidence and the strong performance of the LC model with SARIMA specifications combined with XGBoost for residual modeling, we focus on the quadruple:
\[
\mathbf{Y}_t^{(r)} = \left(\kappa_t^{(m, r)}, \kappa_t^{(f, r)}, \text{ACI}_{t}^{(\text{T10}, r)}, \text{ACI}_{t}^{(\text{T90}, r)} \right),
\]
where
\begin{itemize}
    \item $\kappa_t^{(m, r)}$ is the male period effect at time $t$ in region $r$,
    \item $\kappa_t^{(f, r)}$ is the female period effect at time $t$ in region $r$,
    \item $\text{ACI}_{t}^{(\text{T10}, r)}$ measures the frequency of cold extremes below the 10th percentile, and
    \item $\text{ACI}_{t}^{(\text{T90}, r)}$ measures the frequency of hot extremes above the 90th percentile.
\end{itemize}
Our goal is to build forecasts of $\mathbf{Y}_t^{(r)}$ that retain both the temporal dependence of each component and the cross-sectional dependence among components. In the next subsection, we develop a SARIMA-Copula forecasting approach that achieves this and provides the inputs for mortality projections under Baseline 1 with XGBoost (i.e., the LC-SARIMA model with XGBoost residuals).

\subsection{A SARIMA-Copula forecasting approach} \label{sec:Forecast_Approach}

Having defined the quadruple $\mathbf{Y}_t^{(r)}$, we now specify a joint forecasting method that generates trajectories that respect both the temporal structure of each series and their cross-sectional dependence. A SARIMA model can provide strong marginal forecasts for each component of $\mathbf{Y}_t^{(r)}$, but it treats the four components as independent. In reality, however, it is reasonable to expect that these components are correlated, especially in extreme cases.

To recover the systematic co-movements between mortality period effects and temperature extremes, we adopt a two-layer forecasting approach. The SARIMA-Copula approach consists of:
\begin{enumerate}
    \item \textbf{Marginal forecasting:} apply SARIMA separately to each element of $\mathbf{Y}_t^{(r)}$ to capture monthly seasonality and autocorrelation;  
    \item \textbf{Dependence modeling:} transform residuals from the fitted SARIMA models to uniforms via the probability integral transform and link them with a copula to preserve dependence across $\kappa_t^{(m,r)}$, $\kappa_t^{(f,r)}$, $\text{ACI}_{t}^{(\text{T10}, r)}$, and $\text{ACI}_{t}^{(\text{T90}, r)}$.  
\end{enumerate}

This construction ensures that the mean forecasts are consistent with standard SARIMA models, while the simulated paths reflect realistic joint scenarios. In Sections~\ref{sec:Forecast_SARIMA} and~\ref{sec:Forecast_Copula}, we apply this procedure step by step, first reporting the SARIMA fits and then the copula dependence results for each region.

\subsubsection{Applying the SARIMA model} \label{sec:Forecast_SARIMA}

Let $Y_t$ be an element of $\mathbf{Y}_t^{(r)}$, and $B$ the backward shift operator (i.e., $B(Y_t) = Y_{t-1}$). Similar to \eqref{sarima}, a $\text{SARIMA}(p, d, q)(P, D, Q)_s$ model applied to $Y_t$ can be written as
\[
\left( 1 - \sum_{i=1}^{p} \phi_i B^i \right)
\left( 1 - \sum_{i=1}^{P} \Phi_i B^{is} \right)
(1 - B)^d (1 - B^s)^D Y_t
=
\left( 1 + \sum_{i=1}^{q} \theta_i B^i \right)
\left( 1 + \sum_{i=1}^{Q} \Theta_i B^{is} \right)
\varepsilon_t,
\]
where $\phi_i$ and $\Phi_i$ are the autoregressive and seasonal autoregressive coefficients, $\theta_i$ and $\Theta_i$ are the moving-average and seasonal moving-average coefficients, and $\varepsilon_t$ is a white-noise error with mean zero and variance $\sigma^2$.

Because the observations of $Y_t$ are monthly, the seasonal cycle is set to 12 (i.e., $s=12$). We fit a SARIMA model to each element of $\mathbf{Y}_t^{(r)}$ in all regions $r$, using the \texttt{auto.arima} function in the \texttt{forecast} R package. This function selects the orders and estimates the coefficients automatically based on information criteria. 

Table \ref{tab:FittedSARIMA} reports the fitted SARIMA orders for $\kappa_t^{(m,r)}$, $\kappa_t^{(f,r)}$, $\text{ACI}_{t}^{(\text{T10}, r)}$, and $\text{ACI}_{t}^{(\text{T90}, r)}$ across the six regions. These fitted models will later be used in Section \ref{sec:Forecast_Result} to generate mean forecasts and simulated sample paths beyond the training period.

\begin{table}[h!]
    \centering
    \small
    \renewcommand{\arraystretch}{1.1}
    \begin{tabular}{c|c c c c}
        \toprule
        Region & $\kappa_t^{(f,r)}$ & $\kappa_t^{(m,r)}$ & $\text{ACI}_{t}^{(\text{T90},r)}$ & $\text{ACI}_{t}^{(\text{T10},r)}$ \\
        \midrule
        CEA & (1,0,0)(2,1,0)$_{12}$ & (1,0,1)(2,1,1)$_{12}$ & (1,0,1)(2,1,0)$_{12}$ & (1,0,0)(1,1,0)$_{12}$ \\
        CWP & (3,0,3)(2,1,2)$_{12}$ & (1,0,1)(2,1,1)$_{12}$ & (0,0,1)(2,1,0)$_{12}$ & (2,0,2)(1,1,0)$_{12}$ \\
        MID & (1,0,0)(2,1,0)$_{12}$ & (3,0,0)(2,1,1)$_{12}$ & (2,0,1)(2,1,0)$_{12}$ & (1,0,1)(2,1,1)$_{12}$ \\
        SEA & (2,0,1)(2,1,1)$_{12}$ & (1,0,1)(2,1,1)$_{12}$ & (2,0,2)(2,1,1)$_{12}$ & (0,0,0)(1,1,0)$_{12}$ \\
        SPL & (2,0,2)(2,1,0)$_{12}$ & (2,0,1)(2,1,0)$_{12}$ & (2,0,2)(2,1,2)$_{12}$ & (1,0,1)(2,1,1)$_{12}$ \\
        SWP & (2,0,1)(2,1,0)$_{12}$ & (2,0,2)(1,1,0)$_{12}$ & (2,0,2)(2,1,0)$_{12}$ & (2,0,2)(2,1,2)$_{12}$ \\
        \bottomrule
    \end{tabular}
    \caption{Fitted SARIMA models for $\kappa_t^{(m,r)}$, $\kappa_t^{(f,r)}$, $\text{ACI}_{t}^{(\text{T10},r)}$, and $\text{ACI}_{t}^{(\text{T90},r)}$ across the six regions considered. Orders are reported as $(p,d,q)(P,D,Q)_{12}$.}
    \label{tab:FittedSARIMA}
\end{table}

\subsubsection{Applying the Copula model} \label{sec:Forecast_Copula}

After applying the SARIMA model to capture the autocorrelation and seasonality in the elements of $\mathbf{Y}_t^{(r)}$, we now turn to the underlying dependence structure of $\mathbf{Y}_t^{(r)}$. In particular, we focus on the residuals from the fitted SARIMA models. Let  
\[
\bm{\varepsilon}_t^{(r)} = \big(\varepsilon_t^{(m,r)}, \varepsilon_t^{(f,r)}, \varepsilon_t^{(\text{T10},r)}, \varepsilon_t^{(\text{T90},r)}\big)^\top
\]  
denote the vector of residuals for region $r$. To obtain probability-integral-transformed (PIT) residuals that are suitable for copula modeling, we apply the probability integral transform to each component of $\bm{\varepsilon}_t^{(r)}$, yielding  
\[
\mathbf{U}_t^{(r)} = \big(\hat{F}(\varepsilon_t^{(m,r)}), \hat{F}(\varepsilon_t^{(f,r)}), \hat{F}(\varepsilon_t^{(\text{T10},r)}), \hat{F}(\varepsilon_t^{(\text{T90},r)})\big)^\top,
\]  
where $\hat{F}(\cdot)$ denotes the empirical cumulative distribution function of the residuals.

\begin{figure}[h!]
    \centering
    \includegraphics[width=0.85\textwidth]{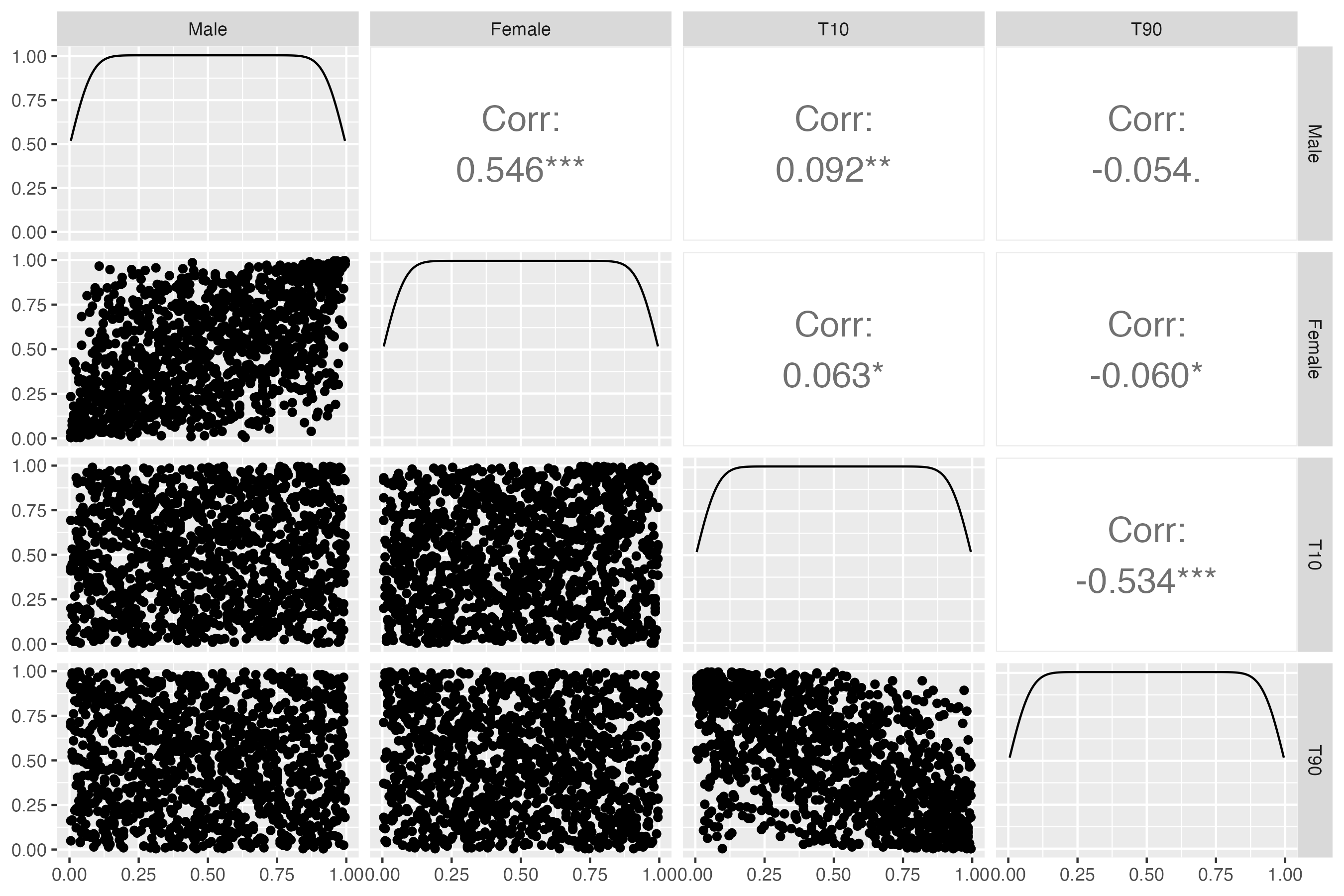}
    \caption{Probability-integral-transformed (PIT) residuals with scatter plots and correlation estimates for $\kappa_t^{(m,r)}$, $\kappa_t^{(f,r)}$, $\text{ACI}_t^{(\text{T10},r)}$, and $\text{ACI}_t^{(\text{T90},r)}$ from all regions.}
    \label{fig:CopulaPITs_All}
\end{figure}

Figure~\ref{fig:CopulaPITs_All} displays the PIT residuals from all regions, shown as scatter plots with correlation estimates. The results reveal substantial dependence between the two period effects, as well as between the temperature extremes and the period effects. In particular, correlations between $\kappa_t^{(m,r)}$ and $\kappa_t^{(f,r)}$ are consistently strong and positive, while negative correlations appear between $\text{ACI}_t^{(\text{T10},r)}$ and $\text{ACI}_t^{(\text{T90},r)}$. Furthermore, both $\kappa_t^{(m,r)}$ and $\kappa_t^{(f,r)}$ remain significantly associated with $\text{ACI}_t^{(\text{T10},r)}$ and $\text{ACI}_t^{(\text{T90},r)}$, supporting the need for a copula framework to capture these cross-dependencies in the joint modeling stage.

To model the cross-dependencies shown, we fit candidate copula families to $\mathbf{U}_t^{(r)}$ using the \texttt{fitCopula} function from the R package \texttt{copula}. Both the Gaussian copula and the Student-$t$ copula with unrestricted correlation matrix are considered. Table~\ref{tab:CopulaFit_AllRegions} reports the AIC and BIC values for each region separately and for the joint sample of all regions. The Student-$t$ copula achieves the lowest AIC and BIC in many cases, including the jointly fitted one. For parsimony, we proceed with the Student-$t$ copula estimated jointly on the residuals from all regions. Estimating a single copula across all regions also improves the efficiency and stability of the dependence parameters.

\begin{table}[h!]
    \centering
    \begin{tabular}{l|cc|cc}
        \toprule
        \multirow{2}{*}{Region} & \multicolumn{2}{c|}{Gaussian Copula} & \multicolumn{2}{c}{Student-$t$ Copula} \\
        \cmidrule{2-5}
         & AIC & BIC & AIC & BIC \\
        \midrule
        CEA & -170.62 & -150.71 & \textbf{-183.58} & \textbf{-160.36} \\
        CWP & \textbf{-52.83} & \textbf{-32.92} & -50.67 & -27.44 \\
        MID & \textbf{-202.83} & \textbf{-182.92} & -200.81 & -177.58 \\
        SEA & -218.06 & \textbf{-198.15} & \textbf{-219.63} & -196.40 \\
        SPL & -159.36 & \textbf{-139.45} & \textbf{-161.43} & -138.21 \\
        SWP & \textbf{-185.39} & \textbf{-165.49} & -184.02 & -160.79 \\
        \midrule
        All Regions & -950.41 & -919.75 & \textbf{-971.83} & \textbf{-936.06} \\
        \bottomrule
    \end{tabular}
    \caption{Fit statistics of Gaussian and Student-$t$ copulas applied to $\mathbf{U}_t^{(r)}$ for each region and for all regions. Lower AIC and BIC values indicate better fit.}
    \label{tab:CopulaFit_AllRegions}
\end{table}

To validate the chosen copula, we compare the empirical correlation matrix of $\mathbf{U}_t^{(r)}$ with the correlation matrix implied by the fitted Student-$t$ copula. Formally, a four-dimensional Student-$t$ copula with correlation matrix $\Sigma$ and degrees of freedom $\nu$ is given by
\[
C_{\Sigma,\nu}(\mathbf{u}) = \mathcal{T}_{\Sigma,\nu}\!\left(\mathcal{T}_\nu^{-1}(u_1), \ldots, \mathcal{T}_\nu^{-1}(u_4)\right),
\]
where $\mathcal{T}_{\Sigma,\nu}$ is the CDF of a multivariate Student-$t$ distribution with parameters $(\Sigma,\nu)$ and $\mathcal{T}_\nu^{-1}$ is the quantile function of a univariate Student-$t$ distribution with $\nu$ degrees of freedom. Table~\ref{tab:CopulaCorrMatrices} reports the empirical and model-implied correlation matrices. The close alignment between the two confirms that the fitted Student-$t$ copula successfully captures the dependence structure, including the strong positive correlation between male and female period effects and the negative dependence between cold and hot extremes.

\begin{table}[h!]
    \centering
    \begin{tabular}{l|cccc|cccc}
        \toprule
        & \multicolumn{4}{c|}{Empirical} & \multicolumn{4}{c}{Model-implied (Student-$t$ Copula)} \\
        \cmidrule{2-9}
        & $\kappa_t^{(m)}$ & $\kappa_t^{(f)}$ & $\text{ACI}_t^{(\text{T10})}$ & $\text{ACI}_t^{(\text{T90})}$
        & $\kappa_t^{(m)}$ & $\kappa_t^{(f)}$ & $\text{ACI}_t^{(\text{T10})}$ & $\text{ACI}_t^{(\text{T90})}$ \\
        \midrule
        $\kappa_t^{(m)}$ & 1.000 & 0.546 & 0.092 & -0.054 & 1.000 & 0.610 & 0.107 & -0.061 \\
        $\kappa_t^{(f)}$ & 0.546 & 1.000 & 0.063 & -0.060 & 0.610 & 1.000 & 0.067 & -0.071 \\
        $\text{ACI}_t^{(\text{T10})}$ & 0.092 & 0.063 & 1.000 & -0.534 & 0.107 & 0.067 & 1.000 & -0.540 \\
        $\text{ACI}_t^{(\text{T90})}$ & -0.054 & -0.060 & -0.534 & 1.000 & -0.061 & -0.071 & -0.540 & 1.000 \\
        \bottomrule
    \end{tabular}
    \caption{Empirical and model-implied correlation matrices of residuals $\mathbf{U}_t^{(r)}$ from all regions.}
    \label{tab:CopulaCorrMatrices}
\end{table}

\subsubsection{Generating joint trajectories of $\mathbf{Y}_t^{(r)}$} \label{sec:Forecast_SimPaths}

Having specified SARIMA models for the marginals and a Student-$t$ copula for the dependence structure, we now combine the two components to generate joint trajectories of $\mathbf{Y}_t^{(r)}$. The SARIMA-Copula framework produces monthly forecasts that capture autocorrelation, seasonality, and cross-sectional dependence, thereby providing realistic inputs for mortality forecasting.

For each region $r$, the following simulation procedure is applied beyond the training period. Let $N$ denote the number of simulated paths and $T$ the forecast horizon in months, with $\tau$ representing the final month of the training period:
\begin{enumerate}
    \item Simulate $N$ realizations of $\mathbf{U}_{\tau+h}^{(r)}$ from the fitted Student-$t$ copula $C^{(r)}_{\Sigma,\nu}$ for $h=1,\ldots,T$.
    \item Transform the simulated $\mathbf{U}_{\tau+h}^{(r)}$ using the empirical quantile functions of the SARIMA residuals to obtain $N$ realizations of $\bm{\varepsilon}_{\tau+h}^{(r)}$.
    \item Apply the fitted SARIMA models with the simulated residuals $\bm{\varepsilon}_{\tau+h}^{(r)}$ to construct $N$ sample paths of $\mathbf{Y}_{\tau+h}^{(r)}$.
    \item Compute mean forecasts and distributional summaries of $\mathbf{Y}_{\tau+h}^{(r)}$ across the $N$ simulated paths.
\end{enumerate}

\begin{figure}[h!]
    \centering
    \includegraphics[width=0.95\textwidth]{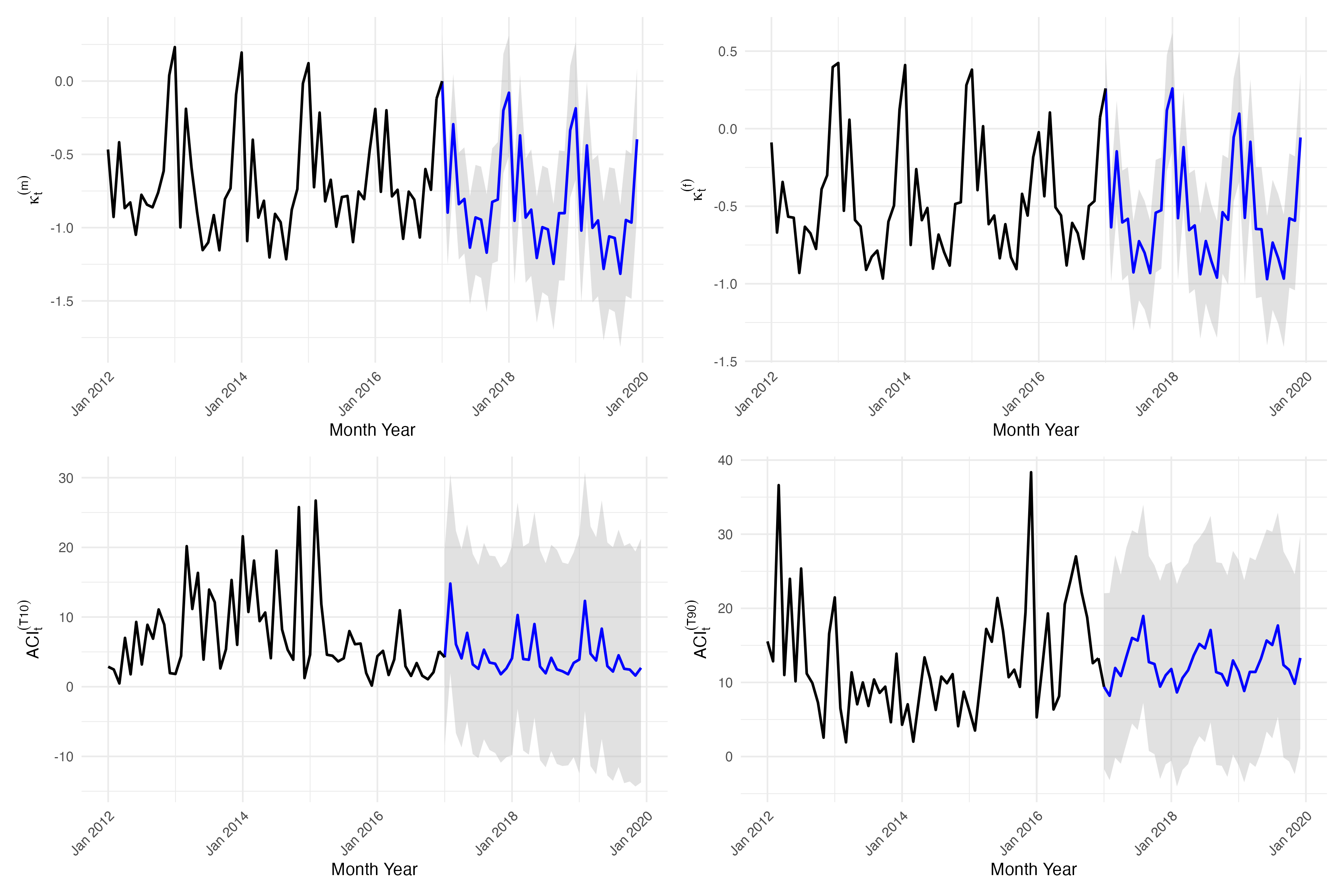}
    \caption{Forecasts of $\kappa_t^{(m,r)}$, $\kappa_t^{(f,r)}$, $\text{ACI}_t^{(\text{T10},r)}$, and $\text{ACI}_t^{(\text{T90},r)}$ for the SEA region using the SARIMA-Copula approach.}
    \label{fig:SEA_Y_Forecast}
\end{figure}

Figure~\ref{fig:SEA_Y_Forecast} displays the SARIMA-Copula forecasts for the SEA region, showing mean trajectories and 95\% prediction intervals for $\kappa_t^{(m,r)}$, $\kappa_t^{(f,r)}$, $\text{ACI}_t^{(\text{T10},r)}$, and $\text{ACI}_t^{(\text{T90},r)}$. The training period is shown in black, with forecasts from 2017 to 2019 plotted in blue and shaded bands. The figure illustrates that the SARIMA-Copula approach preserves the seasonal and cyclical dynamics of each component while producing dependence-aware forecasts that reflect both persistence and variability. These simulated paths provide the inputs for the mortality forecasting models developed in Section~\ref{sec:Forecast_Result}.

\subsection{Forecasting results} \label{sec:Forecast_Result}

We now use the jointly simulated trajectories of $\kappa_t^{(m,r)}$, $\kappa_t^{(f,r)}$, $\text{ACI}_t^{(\text{T10},r)}$, and $\text{ACI}_t^{(\text{T90},r)}$ from the previous subsection to generate future mortality projections. The forecasting framework proceeds in two steps: 
\begin{enumerate}
    \item \textbf{Baseline}: mortality forecasts are obtained by feeding the simulated trajectories of $\kappa_t^{(m,r)}$ and $\kappa_t^{(f,r)}$ into the LC model (equation~\eqref{eq:LCmodel}).
    \item \textbf{Excess}: the baseline forecasts are refined by incorporating the contributions of $\text{ACI}_t^{(\text{T10},r)}$, $\text{ACI}_t^{(\text{T90},r)}$, and other covariates through the XGBoost model, the best-performing specification from Section~\ref{sec:results}.
\end{enumerate}

\subsubsection{Baseline forecasts} \label{sec:LC_baseline}

The first step is to examine the baseline (LC-only) mortality forecasts using jointly simulated trajectories of $\kappa_t^{(m,r)}$ and $\kappa_t^{(f,r)}$. Table~\ref{tab:LC_XGB_ForecastAccuracy} reports RMSE, MAE, and MAPE metrics during the testing period (2017--2019) across the six regions and overall. The results show that the LC model alone provides a reasonable baseline, with overall RMSE around $0.07$ and MAPE near $1\%$. Forecast accuracy is particularly strong in the SEA, SPL, and SWP regions, while performance is relatively weaker in the CWP region.

\begin{figure}[h!]
    \centering
    \includegraphics[width=0.85\textwidth]{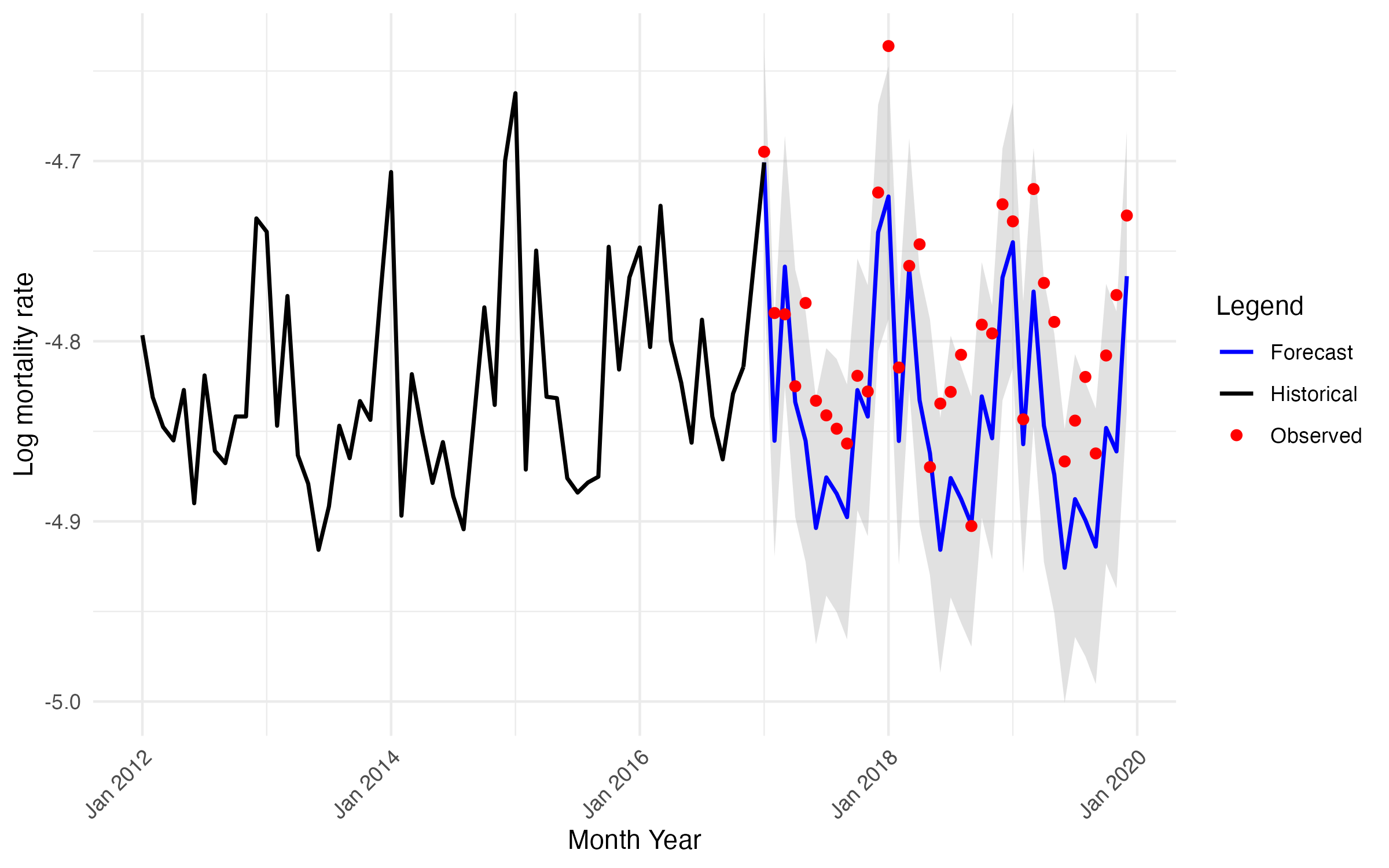}
    \caption{Forecast and observed log mortality rates for females, age group 60–64, from the MID region under the LC baseline model with Copula-SARIMA forecasts.}
    \label{fig:LC_PI_MID}
\end{figure}

To further illustrate forecast performance, Figure~\ref{fig:LC_PI_MID} presents the log mortality rates for females aged 60-64 in the MID region during the testing period. The figure shows the predictive mean and 95\% prediction intervals generated from the LC baseline model using the Copula–SARIMA approach. While the observed mortality rates mostly fall within the forecast bands, the baseline forecasts tend to understate the overall level of mortality and have difficulty capturing extreme high values. This highlights both the usefulness of the baseline model in reflecting seasonal variability and its limitations in anticipating sharp mortality spikes.
 
\subsubsection{Excess forecasts} \label{sec:LC_XGB}

The second step is to improve the baseline mortality forecasts by incorporating climate effec
ts through the XGBoost model. In particular, the projected mortality rate at age $x$ and horizon $t+h$ in region $r$ for gender $g$ is given by
\[
\log \hat{m}_{x,t+h}^{(r,g)} = \log \hat{\mu}_{x,t+h}^{(r,g)} 
+ \hat{R}_{x,t+h}^{(r,g)},
\]
where $\hat{\mu}_{x,t+h}^{(r,g)}$ denotes the baseline LC forecast and $\hat{R}_{x,t+h}^{(r,g)}$ is the predicted residuals from the XGBoost model using age, time, and ACIs as inputs. Age and time are included deterministically, while $\text{ACI}_t^{(\text{T10},r)}$ and $\text{ACI}_t^{(\text{T90},r)}$ are drawn from the simulated Copula-SARIMA trajectories. The remaining ACI covariates are taken from their observed values during the testing period.

Table~\ref{tab:LC_XGB_ForecastAccuracy} compares forecast accuracy between the LC-only baseline and the LC-XGB model across regions. The inclusion of the XGBoost correction systematically improves accuracy, lowering RMSE, MAE, and MAPE in most cases. At the overall level, RMSE falls from $0.0714$ to $0.0633$, and MAPE decreases from $1.14\%$ to $1.01\%$. Notably, the CEA and MID regions show particularly strong gains, with RMSE reductions exceeding $25\%$, while SEA and SPL also demonstrate solid improvements. Overall, the results confirm that incorporating ACIs through XGBoost adds meaningful predictive power beyond the LC baseline.

\begin{table}[h!]
    \centering
    \small
    \renewcommand{\arraystretch}{1.1}
    \begin{tabular}{l | l | c c c}
        \toprule
        Region & Model & RMSE & MAE & MAPE \\
        \midrule
        \multirow{2}{*}{Overall} 
            & LC-only & 0.0710 & 0.0530 & 0.0110 \\
            & LC-XGB  & 0.0629 & 0.0463 & 0.0101 \\
        \midrule
        \multirow{2}{*}{CEA} 
            & LC-only & 0.0710 & 0.0500 & 0.0100 \\
            & LC-XGB  & 0.0542 & 0.0395 & 0.0084 \\
        \multirow{2}{*}{CWP} 
            & LC-only & 0.1000 & 0.0760 & 0.0160 \\
            & LC-XGB  & 0.0947 & 0.0699 & 0.0146 \\
        \multirow{2}{*}{MID} 
            & LC-only & 0.0730 & 0.0560 & 0.0120 \\
            & LC-XGB  & 0.0520 & 0.0402 & 0.0091 \\
        \multirow{2}{*}{SEA} 
            & LC-only & 0.0610 & 0.0480 & 0.0110 \\
            & LC-XGB  & 0.0541 & 0.0421 & 0.0094 \\
        \multirow{2}{*}{SPL} 
            & LC-only & 0.0610 & 0.0470 & 0.0110 \\
            & LC-XGB  & 0.0558 & 0.0426 & 0.0095 \\
        \multirow{2}{*}{SWP} 
            & LC-only & 0.0540 & 0.0410 & 0.0090 \\
            & LC-XGB  & 0.0560 & 0.0436 & 0.0095 \\
        \bottomrule
    \end{tabular}
    \caption{Forecast accuracy of the LC-only model and LC+XGBoost model across six regions and overall, evaluated on the testing period of 2017–2019.}
    \label{tab:LC_XGB_ForecastAccuracy}
\end{table}

\begin{figure}[h!]
    \centering
    \includegraphics[width=0.85\textwidth]{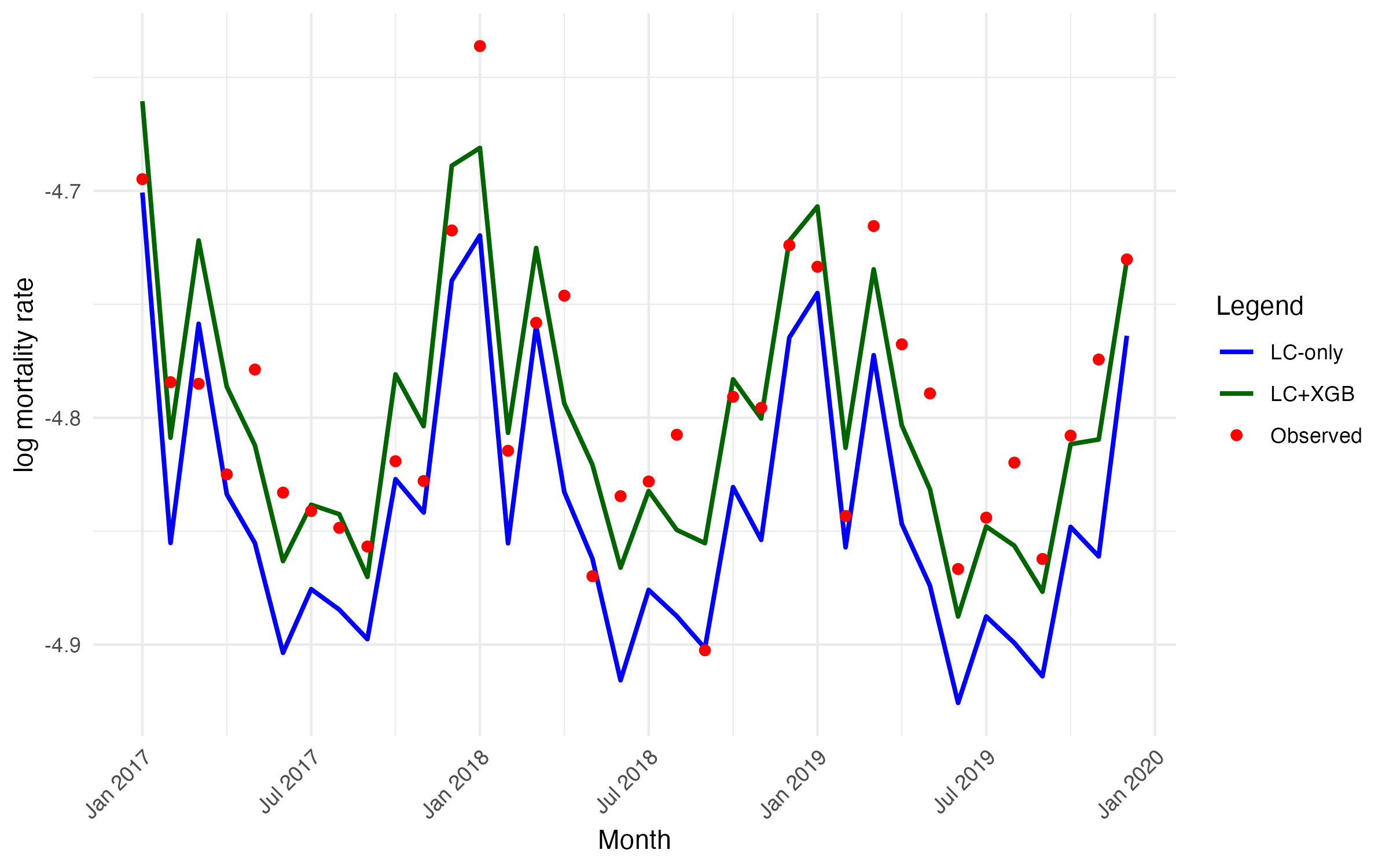}
    \caption{Comparison of forecast means from the LC-only and LC+XGB models for females, age group 60–64, from the MID region during the testing period.}
    \label{fig:LC_XGB_MID}
\end{figure}

Figure~\ref{fig:LC_XGB_MID} compares forecast means from the LC-only and LC+XGB models for females aged 60-64 in the MID region during the testing period. As seen previously, the LC-only model understates the overall level of mortality and fails to capture extreme spikes. By contrast, the LC+XGB model, which augments the baseline with simulated temperature extremes, provides a closer alignment with the observed series. The addition of the fitted XGBoost model reduces the systematic bias and improves responsiveness to short-term fluctuations, addressing the limitations of the LC-only forecasts. This visual comparison complements the accuracy gains reported in Table~\ref{tab:LC_XGB_ForecastAccuracy}, highlighting the value of incorporating climate covariates through machine learning.  

\section{Conclusion} \label{sec:conclusion}

This paper develops and evaluates a unified framework for mortality forecasting that integrates climate variables from the Actuaries Climate Index (ACI) with established stochastic mortality models. The framework consists of two layers: (i) a Lee–Carter baseline with monthly dynamics captured by SARIMA models, and (ii) an excess component linked to ACI covariates through machine learning, where XGBoost achieves the strongest performance. Cross-sectional dependence is modeled using a Student-$t$ copula fitted to the SARIMA residuals, enabling realistic joint simulations of mortality period effects and temperature extremes.

Empirically, the LC-SARIMA baseline produces accurate monthly forecasts of trend and seasonality across U.S. regions. The copula layer enhances forecasting by reproducing the empirical dependence structure, including the strong positive association between male and female period effects and the negative dependence between cold and hot extremes. Incorporating climate covariates through XGBoost yields systematic accuracy improvements, reflected in lower RMSE, MAE, and MAPE values. In summary, the proposed framework with the LC, SARIMA, copula, and XGBoost components forms a unified pipeline that captures both seasonal patterns and climate-sensitive mortality fluctuations.

Our findings have meaningful implications for actuarial practice. By embedding ACI components within a stochastic framework, the model supports mortality projections under extreme temperature conditions, which is relevant for life insurers, pension plans, and public health planning. The baseline–excess structure of the proposed framework secures consistent LC forecasts and augments them with climate-sensitive adjustments through the XGBoost layer. This combination offers methodological soundness and practical flexibility, making the framework applicable to experience studies, scenario analysis, and risk assessment.

There are limitations in our work. First, the framework captures short-term seasonal dependence but not the long-term drift and structural change in mortality-climate interactions. Because both mortality improvement and climate change evolve nonlinearly over decades, a simple SARIMA specification cannot fully capture these dynamics and would require more advanced time-series models (see e.g. \cite{van2016impact} for a structural change model). Second, the copula layer assumes a stationary dependence between mortality period effects and temperature extremes. In reality, this relationship may shift across regimes, for example showing stronger correlations during overlapping heat events and influenza seasons. A dynamic or regime-switching copula could better reflect how climate–mortality coupling evolves over time, for instance in the spirit of the regime-switching model of \cite{robben2025granular3states}.

Overall, our framework demonstrates that integrating stochastic mortality models with data-driven climate indices provides a practical foundation for climate-aware mortality modeling and forecasting. Future extensions could apply this framework to weekly long-horizon projection scenarios with dynamic or regime-switching settings, and explore more advanced machine-learning approaches, such as neural networks, to capture the intrinsic relationships between mortality and climate. These developments would further enhance the framework’s capacity to inform life insurance practice and risk management in an evolving climate environment.

\section*{Acknowledgment}
Kenneth Q. Zhou acknowledges the support of the Natural Sciences and Engineering Research Council of Canada (RGPIN-2025-04157 and DGECR-2025-00488). Karim Barigou acknowledges the support of the Natural Sciences and Engineering Research Council of Canada (RGPIN-2024-04711).

\section*{Data and Code Availability}
The mortality and climate data used in this study are publicly available from the CDC WONDER (\url{https://wonder.cdc.gov}) and the Actuaries Climate Index (\url{https://actuariesclimateindex.org/home/}). Custom R code and cleaned data supporting the results are available from the corresponding author upon reasonable request.

\bibliographystyle{apalike}
\bibliography{reference.bib}

\begin{thebibliography}{}

\bibitem[{American Academy of Actuaries} et~al., 2016]{aci}
{American Academy of Actuaries}, {Canadian Institute of Actuaries}, {Casualty
  Actuarial Society}, and {Society of Actuaries} (2016).
\newblock Actuaries climate index: Development and design.
\newblock Online Resource (accessed on 23 June 2023).

\bibitem[Baccini et~al., 2008]{baccini2008heat}
Baccini, M., Biggeri, A., Accetta, G., Kosatsky, T., Katsouyanni, K., Analitis,
  A., Anderson, H.~R., Bisanti, L., D'Ippoliti, D., Danova, J., et~al. (2008).
\newblock Heat effects on mortality in 15 european cities.
\newblock {\em Epidemiology}, pages 711--719.

\bibitem[Barigou et~al., 2023]{barigou2022bayesian}
Barigou, K., Goffard, P.-O., Loisel, S., and Salhi, Y. (2023).
\newblock Bayesian model averaging for mortality forecasting using
  leave-future-out validation.
\newblock {\em International Journal of Forecasting}, 39(2):674--690.

\bibitem[Boonen and Li, 2017]{boonen2017modeling}
Boonen, T.~J. and Li, H. (2017).
\newblock Modeling and forecasting mortality with economic growth: A
  multipopulation approach.
\newblock {\em Demography}, 54(5):1921--1946.

\bibitem[Cairns et~al., 2006]{cairns2006two}
Cairns, A.~J., Blake, D., and Dowd, K. (2006).
\newblock A two-factor model for stochastic mortality with parameter
  uncertainty: theory and calibration.
\newblock {\em Journal of Risk And Insurance}, 73(4):687--718.

\bibitem[Chen and Guestrin, 2016]{chen2016xgboost}
Chen, T. and Guestrin, C. (2016).
\newblock Xgboost: A scalable tree boosting system.
\newblock In {\em Proceedings of the 22nd acm sigkdd international conference
  on knowledge discovery and data mining}, pages 785--794.

\bibitem[Currie, 2016]{currie2016fitting}
Currie, I.~D. (2016).
\newblock On fitting generalized linear and non-linear models of mortality.
\newblock {\em Scandinavian Actuarial Journal}, 2016(4):356--383.

\bibitem[Curry, 2015]{unitedkingdom}
Curry, C.~L. (2015).
\newblock Extension of the actuaries climate index to the uk and europe: A
  feasibility study.

\bibitem[Garrido et~al., 2023]{garrido:hal-04491982}
Garrido, J., Milhaud, X., and Olympio, A. (2023).
\newblock {The definition of a French actuarial climate index; one more step
  towards a European index}.
\newblock working paper or preprint.

\bibitem[Gasparrini et~al., 2015]{gasparrini2015mortality}
Gasparrini, A., Guo, Y., Hashizume, M., Lavigne, E., Zanobetti, A., Schwartz,
  J., Tobias, A., Tong, S., Rockl{\"o}v, J., Forsberg, B., et~al. (2015).
\newblock Mortality risk attributable to high and low ambient temperature: a
  multicountry observational study.
\newblock {\em The Lancet}, 386(9991):369--375.

\bibitem[Guibert et~al., 2025]{guibert2024impacts}
Guibert, Q., Pincemin, G., and Planchet, F. (2025).
\newblock Impact of climate change on mortality: An extrapolation of
  temperature effects based on time series data in france.
\newblock {\em International Journal of Forecasting}.

\bibitem[Hajat and Kosatky, 2010]{hajat2010heat}
Hajat, S. and Kosatky, T. (2010).
\newblock Heat-related mortality: a review and exploration of heterogeneity.
\newblock {\em Journal of Epidemiology \& Community Health}, 64(9):753--760.

\bibitem[Hanewald, 2011]{hanewald2011explaining}
Hanewald, K. (2011).
\newblock Explaining mortality dynamics: The role of macroeconomic fluctuations
  and cause of death trends.
\newblock {\em North American Actuarial Journal}, 15(2):290--314.

\bibitem[Hyndman and Khandakar, 2008]{hyndman2008automatic}
Hyndman, R.~J. and Khandakar, Y. (2008).
\newblock Automatic time series forecasting: the forecast package for r.
\newblock {\em Journal of Statistical Software}, 27:1--22.

\bibitem[Ishigami et~al., 2008]{ishigami2008ecological}
Ishigami, A., Hajat, S., Kovats, R.~S., Bisanti, L., Rognoni, M., Russo, A.,
  and Paldy, A. (2008).
\newblock An ecological time-series study of heat-related mortality in three
  european cities.
\newblock {\em Environmental Health}, 7(1):1--7.

\bibitem[Lee and Carter, 1992]{lee1992modeling}
Lee, R.~D. and Carter, L.~R. (1992).
\newblock Modeling and forecasting us mortality.
\newblock {\em Journal of The American Statistical Association},
  87(419):659--671.

\bibitem[Li and Tang, 2022]{li2022joint}
Li, H. and Tang, Q. (2022).
\newblock Joint extremes in temperature and mortality: A bivariate pot
  approach.
\newblock {\em North American Actuarial Journal}, 26(1):43--63.

\bibitem[Li and Lee, 2005]{li2005coherent}
Li, N. and Lee, R. (2005).
\newblock Coherent mortality forecasts for a group of populations: An extension
  of the lee-carter method.
\newblock {\em Demography}, 42:575--594.

\bibitem[Liu et~al., 2019]{liu2019statistical}
Liu, Q., Ling, C., and Peng, L. (2019).
\newblock Statistical inference for lee-carter mortality model and
  corresponding forecasts.
\newblock {\em North American Actuarial Journal}, 23(3):335--363.

\bibitem[Ma and Boonen, 2022]{ma2022longevity}
Ma, Q. and Boonen, T.~J. (2022).
\newblock Longevity risk modeling with the consumer price index.
\newblock {\em Available At Ssrn 4209336}.

\bibitem[McMichael et~al., 2008]{mcmichael2008international}
McMichael, A.~J., Wilkinson, P., Kovats, R.~S., Pattenden, S., Hajat, S.,
  Armstrong, B., Vajanapoom, N., Niciu, E.~M., Mahomed, H., Kingkeow, C.,
  et~al. (2008).
\newblock International study of temperature, heat and urban mortality: the
  ‘isothurm’project.
\newblock {\em International Journal of Epidemiology}, 37(5):1121--1131.

\bibitem[Nevruz et~al., 2022]{turkey}
Nevruz, E., Atici, R.~Y., and Yildirak, K. (2022).
\newblock Actuaries climate index: An application for turkey.

\bibitem[Niu and Melenberg, 2014]{niu2014trends}
Niu, G. and Melenberg, B. (2014).
\newblock Trends in mortality decrease and economic growth.
\newblock {\em Demography}, 51(5):1755--1773.

\bibitem[Pan et~al., 2022]{pan2022assessing}
Pan, Q., Porth, L., and Li, H. (2022).
\newblock Assessing the effectiveness of the actuaries climate index for
  estimating the impact of extreme weather on crop yield and insurance
  applications.
\newblock {\em Sustainability}, 14(11):6916.

\bibitem[Plat, 2009]{plat2009stochastic}
Plat, R. (2009).
\newblock On stochastic mortality modeling.
\newblock {\em Insurance: Mathematics And Economics}, 45(3):393--404.

\bibitem[Renshaw and Haberman, 2006]{renshaw2006cohort}
Renshaw, A.~E. and Haberman, S. (2006).
\newblock A cohort-based extension to the lee--carter model for mortality
  reduction factors.
\newblock {\em Insurance: Mathematics And Economics}, 38(3):556--570.

\bibitem[Robben et~al., 2022]{robben2022assessing}
Robben, J., Antonio, K., and Devriendt, S. (2022).
\newblock Assessing the impact of the covid-19 shock on a stochastic
  multi-population mortality model.
\newblock {\em Risks}, 10(2):26.

\bibitem[Robben et~al., 2025a]{robben2024association}
Robben, J., Antonio, K., and Kleinow, T. (2025a).
\newblock The short-term association between environmental variables and
  mortality: evidence from europe.
\newblock {\em Journal of The Royal Statistical Society Series A: Statistics In
  Society}, page qnaf052.

\bibitem[Robben and Barigou, 2025]{robben2025penalized}
Robben, J. and Barigou, K. (2025).
\newblock A penalized distributed lag non-linear lee-carter framework for
  regional weekly mortality forecasting.
\newblock {\em Arxiv Preprint Arxiv:2509.24087}.

\bibitem[Robben et~al., 2025b]{robben2025granular3states}
Robben, J., Barigou, K., and Kleinow, T. (2025b).
\newblock Granular mortality modeling with temperature and epidemic shocks: a
  three-state regime-switching approach.
\newblock {\em Arxiv Preprint Arxiv:2503.04568}.

\bibitem[Seklecka et~al., 2017]{seklecka2017mortality}
Seklecka, M., Pantelous, A.~A., and O'Hare, C. (2017).
\newblock Mortality effects of temperature changes in the united kingdom.
\newblock {\em Journal of Forecasting}, 36(7):824--841.

\bibitem[Serfling, 1963]{serfling1963methods}
Serfling, R.~E. (1963).
\newblock Methods for current statistical analysis of excess
  pneumonia-influenza deaths.
\newblock {\em Public Health Reports}, 78(6):494.

\bibitem[The Institute~of Actuaries, 2023]{australia}
The Institute~of Actuaries, A. (2023).
\newblock Australian actuaries climate index.

\bibitem[Van~Berkum et~al., 2016]{van2016impact}
Van~Berkum, F., Antonio, K., and Vellekoop, M. (2016).
\newblock The impact of multiple structural changes on mortality predictions.
\newblock {\em Scandinavian Actuarial Journal}, 2016(7):581--603.

\bibitem[Villegas et~al., 2018]{andres2018stmomo}
Villegas, A., Kaishev, V.~K., and Millossovich, P. (2018).
\newblock Stmomo: Stochastic mortality modeling in r.
\newblock {\em Journal of Statistical Software}, 84(3):1--38.

\bibitem[Wood, 2017]{wood2017generalized}
Wood, S.~N. (2017).
\newblock {\em Generalized Additive Models: An Introduction with R, Second
  Edition}.
\newblock Chapman And Hall/Crc, 2nd edition.

\bibitem[Xing et~al., 2022]{xing2022projections}
Xing, Q., Sun, Z., Tao, Y., Shang, J., Miao, S., Xiao, C., and Zheng, C.
  (2022).
\newblock Projections of future temperature-related cardiovascular mortality
  under climate change, urbanization and population aging in beijing, china.
\newblock {\em Environment International}, 163:107231.

\bibitem[Yang et~al., 2021]{yang2021projecting}
Yang, J., Zhou, M., Ren, Z., Li, M., Wang, B., Liu, D.~L., Ou, C.-Q., Yin, P.,
  Sun, J., Tong, S., et~al. (2021).
\newblock Projecting heat-related excess mortality under climate change
  scenarios in china.
\newblock {\em Nature Communications}, 12(1):1--11.

\end{thebibliography}

\appendix

\end{document}